\documentclass[12pt]{article}

\usepackage{amssymb}
\usepackage{citesort}
\usepackage{amsmath}
\usepackage{url}

\def\de{\delta}

\def\la{\langle}
\def\ra{\rangle}
\def\pa{\partial}

\def\al{\alpha}

\def\ii{\textrm i}

\newcommand{\db}{de$\,$Broglie}

\newcommand{\dbb}{de$\,$Broglie-Bohm}

\begin{document}
\vspace*{1.0cm}
\noindent
{\bf
{\large
\begin{center}
Pilot-wave theory and quantum fields
\end{center}
}
}

\vspace*{.5cm}
\begin{center}
Ward Struyve{\footnote{Postdoctoral Fellow FWO.}}\\
Institute of Theoretical Physics, K.U.Leuven,\\
Celestijnenlaan 200D, B--3001 Leuven, Belgium.{\footnote{Corresponding address.}}\\
Institute of Philosophy, K.U.Leuven,\\
Kardinaal Mercierplein 2, B--3000 Leuven, Belgium.\\
E--mail: Ward.Struyve@fys.kuleuven.be.
\end{center}

\begin{abstract}
\noindent
Pilot-wave theories provide possible solutions to the measurement problem. In such theories, quantum systems are not only described by the state vector, but also by some additional variables. These additional variables, also called beables, can be particle positions, field configurations, strings, etc. In this paper we focus our attention on pilot-wave theories in which the additional variables are field configurations. The first such theory was proposed by Bohm for the free electromagnetic field. Since Bohm, similar pilot-wave theories have been proposed for other quantum fields. The purpose of this paper is to present an overview and further development of these proposals. We discuss various bosonic quantum field theories such as the Schr{\"o}\-ding\-er field, the free electromagnetic field, scalar quantum electrodynamics and the Abelian Higgs model. In particular, we compare the pilot-wave theories proposed by Bohm and by Valentini for the electromagnetic field, finding that they are equivalent. We further discuss the proposals for fermionic fields by Holland and Valentini. In the case of Holland's model we indicate that further work is required in order to show that the model is capable of reproducing the standard quantum predictions. We also consider a similar model, which does not seem to reproduce the standard quantum predictions. In the case of Valentini's model we point out a problem that seems hard to overcome.
\end{abstract}

\renewcommand{\baselinestretch}{1.1}
\bibliographystyle{unsrt}

\newpage
\tableofcontents

\section{Introduction}
In the pilot-wave theory of \db\ and Bohm for non-relativistic quantum systems \cite{debroglie28,bohm52a,bohm52b}, also known as Bohmian mechanics \cite{durr92}, systems are described by both their wave function and by particle positions. The particles move along deterministic trajectories, under the influence of the wave function. The theory yields the same predictions as standard quantum theory, at least when the latter are unambiguous, when the particles are distributed according to quantum equilibrium.. The incorporation of particles in the description of quantum systems makes it possible to provide an objective description of the world, in which ambiguous notions such as ``measurement'' or ``observer'' play no fundamental role, so that there is no measurement problem. 

The particle positions are often referred to as ``hidden variables'' (like for example in Bohm's seminal papers \cite{bohm52a,bohm52b}). However, as emphasized by Bell \cite[p.\ 201]{bell87a}, this terminology is rather unfortunate, for it is in the particle positions that one finds an image of the visible world. For example, whether Schr\"odinger's cat is alive or dead, or whether a certain outcome is observed in a measurement, depends on the way the particles are configured. The variable that is actually more hidden from us is the wave function, since it only manifests itself by its influence on the particles.

Bell instead preferred to call the particle positions {\em beables} \cite{bell87a}. The beables represent the ontology of the theory, i.e.\ the things which are assumed to exist, independently of being observed in any way.{\footnote{More precisely the particle positions are {\em local beables} \cite{bell76,bell902}, because they relate to particular regions in space-time. In the case of a field ontology the local beables are the fields restricted to space-time regions. The wave function is also a beable but a non-local one. In the terminology of D\"urr {\em et al.}\ the particles constitute the {\em primitive ontology} \cite{durr92,allori06}. This notion is closely related to that of local beables.}} The beables of a theory need of course not be particle positions, they could for example be fields or strings.
 
When it comes to quantum field theory, two possible choices of beables immediately come to mind, namely particle positions and fields. These choices are suggested respectively by the Fock representation and the functional Schr\"odinger representation (where states are functionals on a space of fields) of standard quantum field theory. In this paper we restrict our attention to fields as beables. The first pilot-wave theory with a field configuration as beable was given by Bohm already in 1952 \cite{bohm52b}. In the appendix of his paper he considered this approach for the quantized electromagnetic field. Since Bohm, there have been several proposals for pilot-wave theories for quantum fields where the beables are fields. The purpose of this paper is to present an overview and further development of these proposals. Our attention is focused on the systematic development of the pilot-wave approach, rather than the study of it in various physical situations. For the latter we refer to  \cite{kaloyerou85,kaloyerou94,kaloyerou96,kaloyerou00,kaloyerou03,kaloyerou05,bohm87b,bohm93,holland93b,holland93a,lam941,lam942}.

The first quantum field theory we consider is the bosonically quantized Schr{\"o}\-ding\-er field (which was discussed before in \cite{takabayasi52,holland881,holland93b}). This is a simple field theory which serves as a first illustration of the techniques used in developing a pilot-wave theory. 

Then we consider the free quantized electromagnetic field. First, we show that Bohm's pilot-wave model can be derived by following a general recipe according to which beables are only introduced for gauge independent degrees of freedom. Bohm himself started with imposing the Coulomb gauge. There is also a different approach to gauge theories by Valentini (which was applied to the free electromagnetic field in \cite{valentini92,valentini96}, scalar quantum electrodynamics in \cite{valentini96} and the abelian Higgs model and non-Abelian Yang-Mills theories in \cite{valentini09}). Instead of first identifying the gauge independent degrees of freedom or imposing some gauge fixing, Valentini introduces also dynamics for some gauge degrees of freedom. The actual beable is then considered to be an equivalence class of fields related by the gauge transformations. We will compare the approach of Bohm and that of Valentini in detail for the case of the free electromagnetic field. We find that while the ontologies are strictly speaking different, the theories are empirically indistinguishable, even in quantum non-equilibrium.

After discussing the pilot-wave approach for the free quantized electromagnetic field, we present a pilot-wave model for scalar quantum electrodynamics (a scalar field coupled to the electromagnetic field) and the Abelian Higgs model. In each case we only introduce beables for gauge independent degrees of freedom, thereby avoiding the surplus structure in Valentini's approach. The Abelian Higgs model provides a simple example of a theory that exhibits spontaneous symmetry breaking and the Higgs mechanism. We will see that the pilot-wave approach yields an elegant description of this. 

We also briefly consider non-Abelian Yang-Mills theories, which are the natural generalizations of the electromagnetic field theory. However, due to technical complications we do not provide a pilot-wave model for these theories. The technical complications have to do with the difficulties in identifying and isolating the gauge independent degrees of freedom. We expand on this in a short note.

Then we turn to fermionic quantum field theories. There are two suggestions for introducing field beables for fermionic quantum fields. There is Holland's suggestion \cite{holland881,holland93b} to take a field of angular variables as beable and Valentini's suggestion \cite{valentini92,valentini96} to introduce anti-commuting fields as beables. We discuss in detail the ontology of Holland's model and point out that it is not clear whether the model is capable of reproducing the standard quantum predictions. We also consider an alternative model, which can be developed using the same techniques as in Holland's model. For this model it is simpler to analyse whether or not it reproduces the standard quantum predictions. We find that it seems incapable of doing so. In the case of Valentini's model, we point out a problem that seems hard to overcome. We find that although it is possible to introduce a dynamics for anti-commuting fields, it is unclear what the actual guidance equations and the associated equilibrium probability distribution should be, in order to have a pilot-wave model that reproduces the standard quantum predictions.

Finally, we briefly discuss the pilot-wave models proposed by Struyve and Westman for quantum electrodynamics \cite{struyve06,struyve07c}. The first model is radically minimalist because it contains only beables corresponding to bosonic degrees of freedom, the electromagnetic field degrees of freedom in this case, and none for the fermionic degrees of freedom. The second model is an extension of the first with additional beables for the fermionic degrees of freedom, such as for example a field beable corresponding to mass or charge density. 

In this paper we restrict our attention to field beables. However, we want to emphasize that a field ontology is not the only possible pilot-wave ontology for quantum field theory. One possible alternative is a particle ontology, which seems to be better suited for fermionic field theories. There is for example the seminal work of Bell \cite{bell87b} for lattice quantum field theory where the beables are the numbers of fermions at each lattice point. This work has recently been generalized in two different ways to quantum fields on the continuum. One possible continuum generalization was presented by D\"urr {\em et al.}\ \cite{durr02,durr031,durr032,tumulka03,durr04}. In this approach particles travel deterministically along trajecties, interrupted by stochastic jumps which are usually associated to the creation and annihilation of particles. A different continuum generalization was presented by Colin \cite{colin031,colin032,colin033} for quantum electrodynamics, and generalised by Colin and Struyve \cite{colin07} for other interactions in the standard model. In this approach the Dirac sea is taken seriously, in the sense that particle positions are introduced that correspond to the Dirac sea. A difference between the two models is that the model of Colin is deterministic, whereas the model of D\"urr {\em et al.}\ is, just as Bell's lattice model, stochastic. The fact that there are two generalizations of Bell's lattice model for the continuum originates in a different reading of Bell's work, see e.g.\ \cite{tumulka06,colin07}. 

Before discussing the various pilot-wave models for quantum fields, we will first recall the pilot-wave theory of \db\ and Bohm for non-relativistic quantum systems. Various aspects of this pilot-wave theory, like the explanation of how the standard quantum predictions are reproduced, are straightforwardly generalized to the case of field beables. 

We also outline in general how field beables are introduced. We thereby start with recalling how a quantum field theory can be obtained by quantizing a classical field theory. There are different ways of quantizing a theory, but we focus on a way that is particularly convenient for developing a pilot-wave theory. In particular, it is interesting to see how gauge theories are quantized, since this will also reveal the status of gauge invariance in the corresponding pilot-wave models. For clarity, these quantization methods are standard in quantum field theory and have a priori nothing to do with pilot-wave theory. 

Throughout the paper we use natural units in which $\hbar=c=1$.

\section{From a classical theory to a pilot-wave theory}
There exist several procedures to construct a quantum theory starting from a classical theory, see e.g.\ \cite{twarequeali04} for a recent overview. These procedures are called quantization procedures. Some well-known examples are canonical quantization, path integral quantization and BRST quantization. Even within these different procedures, one has different approaches. The different quantization procedures usually yield, though not always, equivalent quantum theories. 

For the construction of a pilot-wave model for the bosonic quantum field theories we consider in this paper, we found it convenient to start with a formulation of the quantum theory in the context of a particular approach to canonical quantization. This approach consists in identifying the ``unconstrained degrees of freedom'' and by quantizing only these degrees of freedom. Below we make clear what exactly is meant by this. Just to give an idea we can already mention that in the case of gauge theories the unconstrained degrees of freedom are simply the gauge independent degrees of freedom. Beables are then introduced only for some of the unconstrained degrees of freedom.

For fermionic fields the canonical quantization can be carried out in a similar way, but the step of introducing beables causes complications. In this section we will therefore only consider bosonic field theories. The discussion of fermionic field theories is deferred to section \ref{fermionicfields}.

Although our approach works well for the field theories we consider in this paper, it is not necessarily applicable to other field theories. In such cases it could be interesting to look at alternative quantization procedures. Different quantization procedures may further lead to different formulations of a quantum theory, which in turn may suggest the construction of different pilot-wave models. 

In order to illustrate the procedure of canonical quantization, without the technicalities and subtleties that go with quantum field theories, we first consider theories with a finite number of degrees of freedom. After that, we discuss how the transition is made to a field theory. We start with explaining the particular approach to canonical quantization. Then we consider how to construct a pilot-wave model for the obtained quantum theories. We also give a short discussion of the pilot-wave theory originally presented by \db\ and Bohm for non-relativistic quantum systems.

\subsection{Finite number of degrees of freedom}
\subsubsection{Canonical quantization}\label{canonicalquantization}
We first consider theories that concern a finite number of degrees of freedom. We assume that they can be described by the Lagrangian formalism. The Lagrangian $L(q,\dot{q})$ is then a function of some coordinates $q_n$, $n=1,\dots,N$, and the corresponding velocities $\dot{q_n} = d q_n/ d t$. The corresponding Euler-Lagrange equations of motion read
\begin{equation}
\frac{d }{dt} \frac{\partial  L}{\partial  {\dot q}_n } - \frac{\partial  L}{\partial  q_n} = 0\,.
\label{f.1}
\end{equation}

In the canonical quantization approach, the first step is to make the transition from the Lagrangian formulation, where the dynamical variables are the velocity phase-space variables $q_n$ and $\dot{q}_n$, to the Hamiltonian formulation, where the dynamical variables are the momentum phase-space variables $q_n$ and $p_n$. Canonical quantization then proceeds by associating operators to the momentum phase-space variables and by imposing certain commutation relations for these operators. 

In order to arrive at the Hamiltonian formulation we first need the momenta $p_n$ that are canonically conjugate to the coordinates $q_n$. These are defined as
\begin{equation}
p_n = \frac{\partial  L}{\partial {\dot q}_n }(q,\dot{q})\,.
\label{f.2}
\end{equation}
In case these relations can not be inverted to yield the velocities $\dot{q}$ in terms of the coordinates $q$ and the momenta $p$, the Lagrangian is called {\em singular}. The Hamiltonian picture for singular Lagrangians requires some care. The formalism that deals with such Lagrangians was originated by Dirac \cite{dirac64}. Extensive reviews on Dirac's formalism can be found in \cite{dirac64,hanson76,sundermeyer82,gitman90,henneaux91}. Here, we just give some results that are relevant for the paper. In order to keep the presentation simple, we ignore subtleties that are not important for the theories that we will consider here.

The fact that the relations ({\ref{f.2}}) are not invertible implies that there are certain relations between the momentum phase-space coordinates $(q,p)$. These relations can be represented as $\chi_m(q,p)=0$, $m=1,\dots,M'$, and are called {\em constraints}. There may be further constraints that arise from the consistency requirement that the constraints $\chi_m(q,p)=0$ are preserved in time. There is a well-defined algorithm to find these extra constraints, but we do not present it here since the constraints are well known for the theories we will consider. We just assume that we have a complete set of independent constraints $\chi_m(q,p)=0$, $m=1,\dots,M$ (where $M' \leqslant M$). 

One can further define the Hamiltonian function 
\begin{equation}
H = \sum_n p_n \dot{q}_n - L\,,
\label{f.3}
\end{equation}
which is a function of the coordinates $q$ and the velocities $\dot{q}$. By making use of the relations for the momenta ({\ref{f.2}}) and the equations of motion ({\ref{f.1}}), it can be shown that the Hamiltonian can be written as a function $H(q,p)$ of the momentum phase-space coordinates $(q,p)$, despite the fact that some of the velocities $\dot{q}_n$ are not expressible in terms of the coordinates and the momenta. 

For singular systems the equation of motion for a dynamical variable $F(q,p)$ is {\em not} given by the usual relation
\begin{equation}
\dot{F} = [F,H]_P\,,
\label{f.4}
\end{equation}
where $[.,.]_P$ is the Poisson bracket. The reason is that the constraints have to be taken into account. There exists a generalization of the above equation of motion, but for the particular quantization procedure we will consider here, we do not need to go into this.

A constraint $\chi_m$ is called {\em first class} if 
\begin{equation}
[\chi_m,\chi_{m'}]_P\Big|_{\chi_1 = \dots = \chi_M=0}=0\,, \qquad m'=1,\dots,M \,,
\label{f.4.1}
\end{equation}
otherwise it is called {\em second class}. Theories with first class constraints and theories with second class constraints are quantized differently. While one can have theories with both first and second class constraints, we only consider theories which have either of those.

\paragraph{First class constraints}
If the constraints are first class then we are dealing with a gauge theory. A gauge theory is hereby understood as a theory in which the dynamics of some variables is not determined by the equations of motion and the specification of initial data.{\footnote{Although one can adopt alternative notions of gauge theories, this notion is very natural and commonly used for theories whose the equations of motion can be derived from a Lagrangian. This notion of gauge theories further plays an important role in standard methods of quantizing such theories.}} In a theory with first class constraints this underdetermination arises as follows. The equation of motion for a dynamical variable $F$ is given by
\begin{equation}
\dot{F} = [F,H]_P +\sum^M_{m=1}  [F,\chi_m]_P u_m\,,
\label{f.5.1}
\end{equation}
where the $u_m$ are arbitrary phase-space functions which may depend on time, and where the constraints may only be taken into account after the Poisson brackets have been evaluated. If the dynamics of some dynamical variable depends on those arbitrary functions $u_m$, then its time evolution is underdetermined, since different choices of the $u_m$ will lead to different evolutions in time. A map that maps solutions to solutions with the same initial data, as well as a combination of such maps, is called a {\em gauge transformation}. Solutions that are connected by a gauge transformation are called  {\em gauge equivalent}. Dynamical variables for which the evolution does not depend on the arbitrary functions are said to be {\em gauge invariant}. In other words, for gauge invariant variables the time evolution is uniquely determined by the equations of motion and the initial data.  These variables can be regarded as the physical degrees of freedom of the theory. From the equation of motion \eqref{f.5.1} it is clear that gauge independent variables are those for which the Poisson brackets with the constraints are a linear combination of the constraints.

The above definitions also apply in the Lagrangian picture. However, the gauge freedom shows up more explicitly in the Hamiltonian picture. 

Here we consider two ways to quantize a theory with first class constraints:

\begin{itemize}
\item
{\bf Quantizing the gauge independent degrees of freedom:} It can be shown that a canonical transformation can be performed, at least locally, such that the new canonical variables can be written in terms of two sets $q'$ and $q''$, and their respective conjugate momenta $p'$ and $p''$, such that, in terms of the new variables, the constraints read $p''=0$, see e.g.\ \cite[pp.\ 36-45]{gitman90} or \cite{maskawa76}. As such the variables $q'$ and $p'$ can be identified as the gauge independent variables, while the variables $q''$ are gauge variables. 

The equation of motion for a function $F(q',p')$ is generated in the usual way:
\begin{equation}
\dot{F} = [F,H_{ph}]_P\,,
\label{f.6}
\end{equation}
where 
\begin{equation}
H_{ph}(q',p') = H(q',p',q'',p'')\big|_{p''=0}
\label{f.7}
\end{equation}
and where $H(q',p',q'',p'')$ is the Hamiltonian that is obtained from $H(q,p)$ by the canonical transformation. The Hamiltonian $H_{ph}$, which does not depend on the variables $q''$, is called the {\em physical Hamiltonian}. Subsequently, we will generally omit the subscript ``ph'' of the Hamiltonian. The variables $q''$ are the gauge variables; their motion is completely undetermined. Therefore the variables $q''$, together with the variables $p''$ (which are constrained to be zero), can be ignored in the description of the system.

The theory can now be quantized just as in the case of a non-singular Lagrangian. Modulo some potential problems, which we touch upon below, this can be done as follows. First one associates operators ${\widehat q}'$ and ${\widehat p}'$ to the canonical variables $q'$ and $p'$, and imposes the commutation relations
\begin{equation}
[{\widehat q}'_r,{\widehat p}'_s] = \ii \delta_{rs}\,.
\label{f.8}
\end{equation}
Then, using the Schr\"odinger representation 
\begin{equation}
{\widehat q}'_r \to q'_r\,, \qquad {\widehat p}'_r \to - \ii \frac{\partial}{\partial q'_r} \,,
\label{f.8.1}
\end{equation}
where the operators act on square integrable wave functions $\psi(q')$, for which the inner product is given by 
\begin{equation}
\la \psi_1 | \psi_2 \ra = \int dq' \psi^*_1(q')  \psi_2 (q') \,,
\label{f.8.2}
\end{equation}
one obtains the Schr\"odinger equation 
\begin{equation}
\ii\frac{\pa \psi(q',t) }{\pa t} = {\widehat H} \psi(q',t) \,,
\label{f.8.3}
\end{equation}
where ${\widehat H}$ is the operator associated to $H_{ph}$.

This quantization procedure appears very natural, but there might be difficulties in applying it. Although these difficulties do not appear in the theories we consider here, except maybe in non-Abelian Yang-Mills theories, we want to briefly mention some of these here. For more complete discussions see \cite{hanson76,sundermeyer82,gitman90,henneaux91}.

The first difficulty that may arise concerns the identification of the gauge independent degrees of freedom. The problem of finding a canonical transformation which separates the gauge degrees of freedom from the physical degrees of freedom has so far not been constructively solved (except in some simple cases). It is also possible that the canonical transformation can not be performed globally. But for the other theories we consider here, it will always be easy to find a global canonical transformation. 

Secondly, even if the gauge independent degrees of freedom can be identified, it might be that their phase-space is such that the above quantization procedure can not be applied. When the phase-space of gauge independent degrees of freedom, which is parametrized by $q'$ and $p'$, is given by ${\mathbb R}^{2(N-M)}$ there is no problem, but in the case of other phase-spaces it might be necessary to modify the quantization procedure \cite{isham84} (this is for example the case for a particle moving on the positive half-line, where one needs to impose different commutation relations). However, for the theories we consider in this paper we do not need to worry about this; modulo the problems that appear when dealing with an infinite number of degrees of freedom, we can always quantize as outlined above. 

It can further be shown that the physical degrees of freedom $q'$ and $p'$ are unique up to a canonical transformation \cite[p.\ 40]{gitman90}. Choosing to quantize a different set of canonical coordinates which parametrize the physical space does not always lead to an equivalent quantum theory. Possible inequivalence might arise due to the operator ordering ambiguity, i.e.\ equivalent orderings of the classical coordinates might yield inequivalent orderings of the corresponding operators. However, such ambiguities do not concern us here. 

In presenting pilot-wave models for gauge theories, we will always start with presenting the quantum field theory in terms of the gauge independent variables, according to the above framework.

\item
{\bf Imposing constraints as conditions on states:} 
A second way to quantize is to use the quantization rules as in the case of a non-singular Lagrangian
\begin{equation}
[{\widehat q}_r,{\widehat p}_s] = \ii \delta_{rs}
\label{f.9}
\end{equation}
and to impose the constraints as conditions on states 
\begin{equation}
{\widehat \chi_m} |\psi \ra  =0\,, \qquad m=1,\dots,M\,.
\label{f.10}
\end{equation}
The states that satisfy the above condition are called {\em gauge independent} or {\em physical} states. 

This quantization procedure works when the commutators $[{\widehat \chi_{m'}}, {\widehat \chi_{m''}}]$ and $[{\widehat \chi_{m'}}, {\widehat H}]$ are linear combinations of the constraint operators ${\widehat \chi_m}$ (although such relations hold classically, quantum corrections might appear due to quantization \cite{sundermeyer82,henneaux91}). The first condition is required for consistency, the second guarantees that physical states remain physical under the Schr\"odinger evolution. 

A potential difficulty with this approach lies in the definition of the inner product. Working in the Schr\"odinger representation ${\widehat q}_r \to q_r$, ${\widehat p}_r \to - \ii \partial/ \partial q_r$, the inner product 
\begin{equation}
\la \psi_1 | \psi_2 \ra = \int dq \psi^*_1(q)  \psi_2 (q) 
\label{f.10.1}
\end{equation}
will not be finite for physical states. In order to obtain a finite inner product the measure in the integral could be changed, for example by applying the Faddeev-Popov procedure, see e.g.\ \cite[pp.\ 281-283]{henneaux91}. (A simple example is given by the case where one of the constraints reads ${\widehat p}_r  |\psi \ra= 0$. In the Schr\"odinger representation this implies that physical states $\psi(q)$ should not depend on $q_r$, so that the above inner product becomes infinite. However, by inserting the distribution $\delta(q_r)$ in the integral one obtains a finite inner product.) We will discuss this in more detail when considering this quantization procedure for the electromagnetic field.

We mention this way of quantizing, because Valentini's pilot-wave approach fits in this framework.
\end{itemize}

\paragraph{Second class constraints}
Just as in the case of first class constraints, the unconstrained degrees of freedom can be isolated by performing a canonical transformation. It can namely be shown that a canonical transformation can always be performed, at least locally, such that the new canonical variables can be written in terms of two sets $q'$ and $q''$, and their respective conjugate momenta $p'$ and $p''$, such that in terms of the new variables, the constraints read $q''=p''=0$, see \cite[p.\ 27-35]{gitman90} and \cite{maskawa76}. The equation of motion for a function $F(q',p')$ of the unconstrained variables reads
\begin{equation}
\dot{F} = [F,H_{ph}]_P\,,
\label{f.11}
\end{equation}
where the physical Hamiltonian is given by
\begin{equation}
H_{ph}(q',p') = H(q',p',q'',p'')\big|_{q''=p''=0}
\label{f.12}
\end{equation}
and where $H(q',p',q'',p'')$ is the Hamiltonian that is obtained from $H(q,p)$ by the canonical transformation. 

The unconstrained degrees of freedom $q'$ and $p'$ are unique up to a canonical transformation \cite[p.\ 32]{gitman90}.

The resulting theory for the unconstrained canonical variables $q'$ and $p'$ can then be quantized in the usual way.

\paragraph{Some comments on other types of quantization}
In the case of second class constraints the Dirac bracket can be introduced \cite{dirac64,hanson76,sundermeyer82,gitman90,henneaux91}. Unlike the Poisson bracket, the Dirac bracket is consistent with the constraints. This means that it makes no difference whether the constraints are imposed before or after the Dirac bracket is evaluated. An alternative way to quantize then consists in replacing the momentum phase-space coordinates by operators and by imposing commutation relations that are based on the Dirac bracket rather then on the Poisson bracket. However, we do not know how the resulting formulation could be useful for developing a pilot-wave model. Of course one could use the fact that the Dirac bracket can locally be expressed as a Poisson bracket for unconstrained variables, see \cite[p.\ 30]{gitman90} and \cite{maskawa76}, but this then essentially amounts to the quantization procedure for systems with second class constraints we presented above. 

In the case of first class constraints one could also introduce extra constraints, also called {\em gauge constraints}, so that the total set of constraints becomes second class. One could then use techniques employed in the case of second class constraints to quantize. However, again we do not know how to exploit this feature for the purpose of constructing a pilot-wave model. In particular, it can easily be seen that a reformulation of the theory in terms of unconstrained degrees of freedom after gauge fixing, is equivalent to a such reformulation without gauge fixing.

\subsubsection{Construction of a pilot-wave model}\label{construction of a pilot-wave model}
Consider now the construction of a pilot-wave model. For the moment we still consider systems that can be described by a finite number of degrees of freedom. The transition to a field theory is explained in section \ref{fieldtheories}.

Consider a classical Hamiltonian of the form 
\begin{equation}
H= \frac{1}{2} \sum_{r,s} p_r h_{rs} (q) p_s + V(q)\,,
\label{f.14}
\end{equation} 
where $h_{rs}$ is symmetric and where the phase-space coordinates $(q,p) \in {\mathbb{R}}^{2N}$ are unconstrained (see \cite{struyve09a} for the treatment of other Hamiltonians). The classical field Hamiltonians that we will consider in this paper will have a similar form. Quantization proceeds by associating operators ${\widehat q}$ and ${\widehat p}$ to the canonical variables and by imposing the commutation relations
\begin{equation}
[{\widehat q}_r,{\widehat p}_s] = \ii \delta_{rs}\,.
\label{f.15}
\end{equation}
Using the representation 
\begin{equation}
{\widehat q}_r \to q_r \,, \quad    {\widehat p}_r  \to -\ii \frac{\partial}{\partial q_r} \,,
\label{f.16}
\end{equation}
the following Schr{\"o}\-ding\-er equation is obtained
\begin{equation}
\ii\frac{\pa \psi(q,t) }{\pa t} = \left( -\frac{1}{2} \sum_{r,s} \frac{\partial}{\partial q_r}  h_{rs} (q) \frac{\partial}{\partial q_s}  + V(q) \right)\psi(q,t) \,.
\label{f.17}
\end{equation}
The wave function $\psi(q,t)$ can be regarded as the expansion coefficient $\la q| \psi (t)\ra$ of the state $|\psi(t)\ra$ in the basis of states $|q\ra$, where the states $|q\ra$ are eigenstates of the operators ${\widehat q}_r$, i.e.\ ${\widehat q}_r|q\ra = q_r |q\ra$.

There is an operator ordering ambiguity in the kinetic term (for example, regarding $h_{rs}$ as a metric, the Laplace-Beltrami operator could be introduced instead). However, the operator ordering that is used here is the most natural for the theories we will consider. With this operator ordering, the Hamiltonian operator is also Hermitian with respect to the inner product
\begin{equation}
\la \psi_1 | \psi_2 \ra = \int dq \psi^*_1(q)  \psi_2 (q) \,.
\label{f.18}
\end{equation}
Other operator orderings could potentially lead to different quantum theories.

According to standard quantum theory, $|\psi(q,t)|^2 dq=|\la q| \psi (t)\ra|^2dq$ yields the probability that a measurement of the observable ${\widehat q}$ yields an outcome which is in a volume element $dq$ around the configuration $q$. The Schr{\"o}\-ding\-er equation implies the following continuity equation for the density $|\psi(q,t)|^2$:
\begin{equation}
\frac{\pa |\psi|^2 }{\pa t}  + \sum_r \frac{\partial j^\psi_r}{\partial q_r}  =0\,,
\label{f.19}
\end{equation}
where 
\begin{equation}
j^\psi_r = \frac{1}{2\ii} \sum_s  h_{rs} \left( \psi^* \frac{\partial \psi}{\partial q_s}   - \psi    \frac{\partial \psi^*}{\partial q_s}   \right)  = |\psi|^2  \sum_s  h_{rs} \frac{\partial S}{\partial q_s}  
\label{f.20}
\end{equation}
is the probability current, with $\psi=|\psi|e^{\ii S}$.

A pilot-wave model can be constructed by introducing a beable configuration $q$, which takes values in the configuration space ${\mathbb R}^N$, and which evolves in time according to the guidance equation
\begin{equation}
\dot{q}_r = \frac{j^\psi_r}{|\psi|^2} =  \sum_s  h_{rs} \frac{\partial S}{\partial q_s} \,.
\label{f.20.1}
\end{equation}
This dynamics implies that if the beable configurations are distributed according to $|\psi(q,t_0)|^2$ over an ensemble at a certain time $t_0$, then they are distributed according to $|\psi(q,t)|^2$ at other times $t$. This means that the distribution $|\psi|^2$ keeps its functional form of the wave function $\psi$. This property is called {\em equivariance} \cite{durr92}. The distribution $|\psi|^2$ plays the role of an equilibrium distribution \cite{bohm53b,valentini91a,valentini92,durr92,durr09,bohm93}, similar to that of thermal equilibrium in classical statistical mechanics, and is called the {\em quantum equilibrium distribution}. In this paper we always assume quantum equilibrium. 

In some models some additional beables will be introduced by means of Holland's local expectation value \cite{holland93b}. The local expectation value corresponding to a Hermitian operator ${\widehat Q}$ and a wave function $\psi(q,t)$ is the function 
\begin{equation}
{\mathcal{Q}}(q,t) = {\textrm{Re}} \frac{\int dq' \psi^*(q,t) \langle q |{\widehat Q}| q' \rangle \psi(q',t)}{ |\psi(q,t)|^2} \,.
\label{f.20.101}
\end{equation}
A new beable $Q$ can then be introduced, by evaluating the local expectation value for the actual value of the beable configuration $q$, i.e.
\begin{equation}
Q(t) = {\mathcal{Q}}(q(t),t) \,.
\label{f.20.102}
\end{equation}
As such its time evolution is completely determined by the time evolution of the wave function and that of the beable configuration. In quantum equilibrium, the distribution of the beable $Q$ is given by
\begin{equation}
\rho(Q,t) = \int dq |\psi(q,t)|^2  \delta(Q - {\mathcal{Q}}(q,t)) \,.
\label{f.20.103}
\end{equation}
This distribution is in general different from the distribution $|\langle Q | \psi \rangle|^2$, where the states $|Q\rangle$ are the eigenstates of ${\widehat Q}$. The definition by means of Holland's local expectation value only guarantees that the expectation value of the beable $Q$ agrees with the quantum mechanical expectation value $\langle \psi |{\widehat Q}| \psi \rangle$. However, during actual measurements it might be that the beable $Q$ becomes distributed according the quantum mechanical distribution. This happens for example for such beables introduced in the context of the non-relativistic pilot-wave theory of \db\ and Bohm, see \cite[pp.\ 339-347]{holland93b}. Note that we definitely do not want to introduce a beable for every Hermitian operator in this way. In most cases the introduction of such additional beables is even unnecessary.

We want to end this section with a word of caution concerning the choice of beable. The beables $q_r$ (and $Q$) that we introduced correspond to quantum operators ${\widehat q}_r$ (and ${\widehat Q}$) and one could get the impression that we could have chosen any operator for which to introduce beables. However, this is not correct. The beables should yield an image of the visible world (i.e.\ they should account for things like tables, chairs, cats, instrument pointers, etc.). Therefore, the beables should in the first place relate to an ontology in space-time. If the beables are for example associated to momentum operators or spin operators, without a specification of how these beables relate to, or influence the behaviour of, some ontology in space-time, it is unclear how they could yield an image of the visible world \cite{tumulka062}. Here we are concerned with models in which the beables are fields in space and these could potentially yield an image of the visible world.

\subsubsection{Example: Pilot-wave theory of de$\,$Broglie and Bohm for non-relativistic quantum systems}\label{examplepilotwaveinterpretationnonrelativistic}
In this pilot-wave theory \cite{debroglie28,bohm52a,bohm52b}, the Schr{\"o}\-ding\-er equation is given by 
\begin{equation}
\ii  \frac{\partial \psi({\bf x}_1, \dots,{\bf x}_n,t)}{\pa t} =  \left( -\sum^n_{k=1} \frac{ \nabla^2_k}{2m_k} + V({\bf x}_1, \dots,{\bf x}_n) \right)  \psi({\bf x}_1, \dots,{\bf x}_n,t)\,,
\label{f.20.2}
\end{equation}
where $\psi({\bf x}_1,\dots,{\bf x}_n,t)$ is, for all times $t$, a function on configuration space ${\mathbb R}^{3n}$. The beables are $n$ particle positions in physical space ${\mathbb R}^3$, denoted by ${\bf x}_k$, $k=1,\dots,n$, for which the evolution equation is given by 
\begin{equation}
\dot{ {\bf x}}_k =  \frac{1}{2\ii m_k|\psi|^2}\left(\psi^* {\boldsymbol {\nabla}}_k \psi - \psi  {\boldsymbol {\nabla}}_k \psi^* \right) = \frac{{\boldsymbol{\nabla}}_k S}{m_k}  \,.
\label{f.20.3}
\end{equation}
The quantum equilibrium distribution is given by $|\psi({\bf x}_1,\dots,{\bf x}_n,t)|^2$. 

Bohm actually presented the dynamics in a Newtonian form, with an extra $\psi$-dependent potential, called the {\em quantum potential}, thereby regarding the guidance equation as an extra constraint on the possible momenta. However, while this Newtonian formulation has some applications, it tends to obscure and complicate things. Therefore, we will consider only formulations in terms of guidance equations throughout this paper.

Having the quantum equilibrium distribution for ensembles is a key ingredient in showing that the pilot-wave theory reproduces the predictions of standard quantum theory. As we will explain below, it guarantees that the Born rule is recovered. As shown by D\"urr {\em et al.}\ \cite{durr92}, most initial configurations of the universe (relative to the natural measure $|\Psi(X)|^2dX$, with $\Psi$ the wave function of the universe) yield the quantum equilibrium distribution $|\psi(x)|^2$ for actual ensembles described by the effective wave function $\psi(x)$. As such, one also expects that most non-equilibrium distributions tend to evolve to equilibrium. This was indeed illustrated by numerical simulations \cite{valentini05,colin10}. 

Another key ingredient of the pilot-wave theory is that wave functions representing macroscopically distinct states (like for example a cat in the alive state or dead state, or macroscopically different orientations of an instrument needle) have their support approximately concentrated in different regions in configuration space. That is, these wave functions have their support approximately concentrated on configurations that on the macroscopical level yield the same familiar image of the macroscopic objects. As a result, in quantum equilibrium, the particles will typically attain one of those configurations and as such yield an image of the object. Even if the wave function happens to be in a superposition of macroscopically distinct states of a macroscopic object, the configurations will typically end up displaying either state of the object. For example in the case of Schr\"odinger's cat, where the wave function is in a superposition of the state of a live cat and a dead cat, the configurations will typically end up either displaying a live cat or a dead cat. In particular, in a measurement situation, the non-overlap of wave functions representing macroscopically distinct states of macroscopic pointers, like for example instrument needles, will guarantee that measurement results get recorded and displayed in the position configurations of those pointers. In addition, it will also give rise to an effective collapse. 

Let us now consider the pilot-wave description of a measurement situation in detail and see how the standard quantum predictions are reproduced. This analysis was first given in \cite{bohm52b} and has been repeated many times afterwards, see e.g.\ \cite{bell87b,bohm87a} (for an extensive analysis see \cite{durr033}). In standard quantum theory a measurement situation can be described as follows. Before the measurement the wave function is given by a product $\psi^s \psi^a$, where $\psi^s=\sum_i c_i \psi^s_i$ is the wave function of the system, with $\psi^s_i$ the different possible eigenstates of the operator that is being measured, and $\psi^a$ is the wave function of the macroscopic pointer in the ready state (one could also include the rest of the relevant environment in $\psi^a$). The macroscopic pointer could for example be an instrument needle. During the measurement process, the wave function evolves to the entangled form $\sum_i c_i \psi^s_i \psi^a_i$. The $\psi^a_i$ are the different states of the macroscopic pointer corresponding to the different possible outcomes of the measurement (for example an instrument needle pointing in different directions). The wave function then collapses to one of the terms $\psi^s_i \psi^a_i$, say $\psi^s_k \psi^a_k$, and this with probability $|c_k |^2$. The wave function of the system ends up in the $k$-th eigenstate of the operator and the wave function of the pointer represents the $k$-th outcome of the measurement.

In the pilot-wave theory the measurement situation is described as follows. As in standard quantum theory the wave function evolves into a superposition $\sum_i c_i \psi^s_i \psi^a_i$, but this time there is no subsequent collapse. The measurement result does not get recorded in the wave function, but in the configuration $({\bf x}^a_1,\dots,{\bf x}^a_n)$ of particle positions of the pointer. This can be seen as follows. Each $\psi^a_i$ has its support approximately concentrated on configurations that display a particular macroscopic state of the pointer. This implies that the $\psi^a_i$ have negligible overlap and that the beable configuration $({\bf x}^a_1,\dots,{\bf x}^a_n)$ will be in the support of only one of the wave functions $\psi^a_i$, say $\psi^a_k$. As such it displays the $k$th macroscopic state of the pointer and hence the corresponding measurement outcome. Since in quantum equilibrium, the probability for the configuration $({\bf x}^a_1,\dots,{\bf x}^a_n)$ to be in the support of $\psi^a_k$ is given by $|c_k |^2$, also the quantum probabilities are recovered. Note that the wave functions $\psi^s_i$ might be overlapping in which case the configuration of the system itself does not display the measurement outcome. However since measurement results are revealed to us be means of macroscopic records, this does not pose a problem. 

Further, if the different wave functions $\psi^s_i \psi^a_i$ stay approximately non-over\-lap\-ping at later times, which is usually guaranteed by decoherence effects, then as one can easily verify from the guidance equations, only one of the wave functions, namely $\psi^s_k \psi^a_k$, will play a significant role in determining the velocity field of the actual beable configuration of system and apparatus. Since the other wave functions $\psi^s_i \psi^a_i$, $i \neq k$ play only a negligible role, they can be ignored in the future description of the behaviour of the beable configuration. This is called an effective collapse. The effective collapse explains the success of the ordinary collapse in standard quantum theory. Unlike the collapse in standard quantum theory the effective collapse is not one of the axioms of the theory, but a consequence of the theory.{\footnote{In the \dbb\ theory, the wave function of a subsystem of the universe can be defined as the {\em conditional wave function} \cite{durr92}. This wave function actually undergoes what could be called an actual collapse.}}

Note that the property that wave functions of macroscopically different states are (approximately) non-overlapping in configuration space is a special property of the position representation. The same wave functions might not be non-overlapping in an other representation. This is very important for the choice of beable since it is unclear how a pilot-wave model in which such wave functions are overlapping in the corresponding configuration space may reproduce the quantum predictions. Similar issues will arise when introducing field beables for quantum field theory, i.e.\ merely introducing some field beable which is distributed according to a quantum equilibrium distribution will not guarantee that the pilot-wave model reproduces the standard quantum predictions. In section \ref{alternativemodel} a non-trivial example of such a theory is considered.

\subsection{Field theories}\label{fieldtheories}
So far we have only considered systems that can be described by a finite number of degrees of freedom. The transition to a system that is described by a continuum of degrees of freedom is straightforward. The transition can be thought of as a replacement of the discrete label $n$ of the coordinates $q_n$ by a continuum label ${\bf x}$, i.e.\ $q_n \to  \phi({\bf x})$. The sums over the discrete label $n$ are then replaced by integrals over ${\bf x}$ and the derivatives $\partial / \partial q_n$ are replaced by functional derivatives $\delta / \delta \phi({\bf x})$. The corresponding field velocities $\partial \phi(t,{\bf x})/ \partial t$ will be denoted as ${\dot \phi}(t,{\bf x})$. The field momenta $\delta L/ \delta {\dot \phi}({\bf x})$ will be denoted as $\Pi_{\phi}({\bf x}) $. Usually we deal with a number of fields so that the fields also carry an extra discrete label $r$, i.e.\ $\phi_r({\bf x})$.

The Hamiltonian formulation of a field theory involves some subtleties that are not present for a system that can be described by a finite number of degrees of freedom. However these subtleties are not important for the field theories we consider here. The interested reader is referred to \cite{hanson76,sundermeyer82,gitman90,henneaux91}.

\subsubsection{Introducing field beables}\label{introducingfieldbeables}
For the field theories we consider here, the constraints are of the form $\chi_m(\phi({\bf x}),\Pi_{\phi}({\bf x}))$, $m=1,\dots,M$. So for each $m$ there is an infinite number of constraints, one for each point ${\bf x}$ in space (however, for convenience, we will often say that we have $M$ constraints, instead of saying that we have an infinite number of them). The physical Hamiltonians will be of the form
\begin{equation}
H= \frac{1}{2} \sum_{r,s} \int d^3 x  d^3y  \Pi_{\phi_r}({\bf x}) h^\phi_{rs}({\bf x},{\bf y})  \Pi_{\phi_s}({\bf y}) + V(\phi)\,,
\label{f.21}
\end{equation} 
where the superscript in $h^\phi_{rs}({\bf x},{\bf y})$ denotes that this density can be a functional of $\phi$. 

On a formal level, the corresponding quantum field theory is obtained as follows. The canonical variables are replaced by operators and the commutation relations 
\begin{equation}
[{\widehat \phi}_r({\bf x}),{\widehat \Pi}_{\phi_s}({\bf y})] = \ii\delta_{rs}\delta({\bf x}-{\bf y})
\label{f.22}
\end{equation}
are imposed. These commutation relations are realized by the representation
\begin{equation}
{\widehat \phi}_r({\bf x})\to \phi_r({\bf x})\,, \quad {\widehat \Pi}_{\phi_r}({\bf x}) \to -\ii \frac{\delta }{\delta \phi_r({\bf x})}\,.
\label{f.23}
\end{equation}
These operators act on wave functionals $\Psi(\phi)$, for which the inner product is given by
\begin{equation}
\la \Psi_1 | \Psi_2 \ra =\int \left( \prod_r {\mathcal D} \phi_r \right) \Psi^*_1(\phi) \Psi_2(\phi)  \,,
\label{f.24}
\end{equation}
where ${\mathcal D} \phi_r =\prod_{{\bf x}} d \phi_r({\bf x})$. By introducing the field basis $|\phi\ra$, where the $|\phi\ra$ are eigenstates of the operators ${\widehat \phi}_r$, i.e.\ ${\widehat \phi}_r({\bf x})|\phi\ra = \phi_r({\bf x}) |\phi\ra$, the wave functional $\Psi(\phi,t)$ can be regarded as the expansion coefficient $\la \phi| \Psi (t) \ra$ of the state $|\Psi(t)\ra$. The dynamics of the wave functional is determined by the functional Schr{\"o}\-ding\-er equation
\begin{equation}
\ii\frac{\pa \Psi(\phi,t)}{\pa t} = \left( - \frac{1}{2} \sum_{r,s} \int d^3 x  d^3y  \frac{\delta }{\delta \phi_r({\bf x})} h^\phi_{rs}({\bf x},{\bf y})  \frac{\delta }{\delta \phi_s({\bf y})}  + V(\phi)\right)\Psi(\phi,t) \,.
\label{f.25}
\end{equation}

The representation for the field operators is the functional Schr{\"o}\-ding\-er representation. It is the natural generalization of the familiar representation (\ref{f.16}) for ordinary quantum mechanics, but is not so widely used as for example the Fock representation. An introduction to the functional Schr{\"o}\-ding\-er picture can be found in \cite{jackiw95}. See also \cite{hatfield91} for a detailed treatment of quantum field theory in the functional Schr{\"o}\-ding\-er picture. 

A pilot-wave model can be constructed by introducing field beables $\phi_r({\bf x})$ whose dynamics is determined by the guidance equations
\begin{equation}
{\dot \phi}_r({\bf x}) = \sum_s \int d^3 y h^\phi_{rs}({\bf x},{\bf y})\frac{\delta S}{\delta \phi_s({\bf y})} \,,
\label{f.26}
\end{equation}
where $\Psi = |\Psi|e^{\ii S}$ was used. The quantum equilibrium distribution is given by $|\Psi(\phi,t)|^2 \prod_r {\mathcal D} \phi_r$.

\subsubsection{Regularization, renormalization and probability interpretation}\label{regularizationandrenormalization}
The presentation of the pilot-wave model was merely formal, lacking mathematical rigour. In particular, the measure $\prod_r {\mathcal D} \phi_r$ that appears in the inner product \eqref{f.24} and in the quantum equilibrium distribution is ill-defined. It was treated as an infinite-dimensional generalization of the Lebesgue measure, but such measures do not exist. Instead, one should adopt the infinite-dimensional analogue of a weighted measure \cite{glimm87,isham91,isham92}. With such a measure, the functional Schr\"odinger representation \eqref{f.23} needs to be modified in order to ensure that the operators ${\widehat \Pi}_{\phi_r}$ are Hermitian. In addition, care is required in the formulation of the Schr\"odinger equation, since such a measure is generally concentrated on distributions that are not smooth functions \cite{colella73}. 

Another approach to make the Schr\"odinger picture well-defined was presented in \cite{symanzik81,luscher85,cooper87,pi87,luscher92}. There it was shown, at least for specific theories, that the functional Schr{\"o}\-ding\-er picture can be made well-defined by a suitable regularization and renormalization. 

Given a well-defined functional Schr\"odinger equation, we also expect to be able to find well-defined guidance equations.

The difficulties of course disappear when a regularization is introduced that makes the number of degrees of freedom finite. One such regularization consists in assuming a finite spatial volume, together with some appropriate boundary conditions, and an ultra-violet momentum cut-off. The assumption of a finite spatial volume discretizes the possible momenta. The cut-off then further makes the total number of possible momenta finite. More explicitly, these assumptions imply that the fields $\phi_r({\bf x})$ have Fourier expansions
\begin{equation}
\phi_r({\bf x}) = \frac{1}{\sqrt{L^3}} \sum_{ {\bf k} } e^{\ii {\bf k} \cdot {\bf x}} \phi_{r{\bf k}} \,,
\label{f.27}
\end{equation}
where the possible momenta ${\bf k}$ are given by $(n_1,n_2,n_3)2\pi/L$, with the $n_i \in {\mathbb N}$ such that $|{\bf k}|\leqslant \Lambda$. $L$ is the length of a side of the cubic spatial volume and $\Lambda$ is the cut-off. In this way, the theory can be formulated in terms of the variables $\phi_{r{\bf k}}$, of which there are finitely many. The Schr{\"o}\-ding\-er equation will then be an ordinary wave equation for wave functions $\psi(\phi_{r{\bf k}_1},\dots,\phi_{r{\bf k}_n},t)$ and the beables will be a set of variables $\phi_{r{\bf k}_i}$, $i=1,\dots,n$, which are distributed according to $|\psi(\phi_{r{\bf k}_1},\dots,\phi_{r{\bf k}_n},t)|^2$ in quantum equilibrium. The beables $\phi_{r{\bf k}_i}$ will still give rise to a field $\phi_r({\bf x})$ in physical space, through equation (\ref{f.27}). 

This regularization makes both the Schr{\"o}\-ding\-er equation and the guidance equation cut-off dependent. Removing the cut-off, by applying a suitable renormalization scheme and taking the cut-off to infinity is in general very hard (already on the level of the wave equation). On the other hand, it may be that nature provides a natural cut-off, so that the limit process is not required. 

Another possible regularization which makes the total number of degrees of freedom finite consists in assuming a finite spatial lattice. The degrees of freedom are then the ``field'' values at each lattice point. This type of regularization is perhaps more elegant than the cut-off regularization, since lattice gauge theories can be formulated in way which makes the gauge structure very similar to the gauge structure of the continuum theory (see for example the Kogut-Susskind approach \cite{kogut75,kogut79}, in which space is treated as discrete and time as continuous). 

In case the regulators are not removed by some renormalization scheme, the pilot-wave ontology in the case of a lattice regularization would be different from the one in the case of the cut-off regularization that was described above (even though the lattice spacing actually also yields an ultra-violet cut-off, which is of the order of the inverse of the lattice spacing). In the case of a lattice regularization, the beables would represent degrees of freedom at the lattice points, whereas with the cut-off regularization the beables would be fields on space. 

In this paper, we maintain the formal approach, i.e.\ the approach outlined in the previous section, in developing the pilot-wave models. Although we have not studied in detail the effects of a more rigorous approach, we do not expect major deviations from the more formal formulation.

\subsubsection{Reproducing the quantum predictions}\label{reproducingquantumpredictions}
As mentioned before in the context of the non-relativistic pilot-wave theory of \db\ and Bohm, cf.\ section \ref{examplepilotwaveinterpretationnonrelativistic}, assuming some beable that is distributed according to quantum equilibrium is not sufficient to guarantee that a pilot-wave model reproduces the predictions of standard quantum theory. The beables should yield, on the macroscopic level, an image of macroscopic objects. Therefore, it is important to consider wave functionals representing macroscopic states and find out what the typical field configurations are in quantum equilibrium. In particular, wave functionals representing different macroscopic states should have their support approximately concentrated on different regions in field space. For example, whether Schr\"odinger's cat is alive or dead, or whether there is a cat in the first place, depends on the values the field beables take and if the wave functionals representing a live cat and a dead cat would be very much overlapping, then there would be nothing in the field beable revealing whether we are dealing with a live cat or a dead cat. Similarly, in measurement situations, the beables should display a definite state of the macroscopic pointer. If the wave functionals corresponding to different states of the macroscopic pointer have significant overlap then there is no hope that the pilot-wave model wil reproduce the standard quantum predictions. 

In \cite{saunders99} Saunders expressed a worry for using fields as beables. He expressed some doubts whether localized macroscopic bodies are represented by localized field beables. If not, he claimed, it would be unclear how a pilot-wave model with field beables reproduces the quantum predictions. However, although it is true that a pilot-wave model in which localized macroscopic bodies are represented by localized fields may reproduce the quantum predictions, it is by no means a necessary requirement. Macroscopic bodies could also be displayed by non-localized fields. For example macroscopic objects could be represented by fields that differ in magnitude in certain region of physical space. But although a field configuration representing a localized macroscopic object need not be localized, it is still desired that the field configuration displays the image of localized macroscopic object {\em locally}, i.e.\ it is desired that the presence of the macroscopic object can be inferred from the characteristics of the field configuration in the spatial region where the object is supposed to be located. Although an ontology which does not satisfy this property can possibly be maintained in some cases, it would be rather far removed from our everyday experience of the macroscopic world.

\subsubsection{Non-locality and Lorentz invariance}
We want to end this section with a note on non-locality and Lorentz invariance. The pilot-wave models we present here are non-local. Because of Bell's theorem the non-locality is unavoidable. The pilot-wave models are further formulated with respect to a preferred frame of reference. In the context of a relativistic quantum field theory this will imply pilot-wave models that are not Lorentz invariant. Nevertheless, in quantum quilibrium these pilot-wave models will reproduce the standard quantum theoretical predictions. Therefore the empirical predictions of the pilot-wave models will be Lorentz invariant, at least when the quantum theoretical predictions are Lorentz invariant \cite{bohm93,holland93b,berndl96,durr99}. There are some attempts to formulate Lorentz covariant pilot-wave models. An overview of these is given by Tumulka \cite{tumulka06}. See however also Valentini \cite{valentini97} for an argument that the fundamental symmetry of pilot-wave theories is Aristotelian rather than Lorentzian.

\section{Schr{\"o}\-ding\-er field}\label{schroedingerfield}
The first field theory that is considered is the quantized non-relativistic free Schr{\"o}\-ding\-er field. The Schr{\"o}\-ding\-er field can be quantized by imposing either bosonic or fermionic commutation rules. Here we only discuss bosonic quantization. Fermionic quantum field theories will be discussed in section \ref{fermionicfields}.

The pilot-wave approach to the bosonically quantized Schr{\"o}\-ding\-er field was discussed before by Takabayasi \cite{takabayasi52} and Holland \cite{holland881} and \cite[pp.\ 449-451]{holland93b}. Our presentation differs from that of Takabayasi and Holland. We also obtain a slightly different pilot-wave model.  

\subsection{Pilot-wave model}
First consider the quantization of the Schr{\"o}\-ding\-er field. The classical field equation is just the non-relativistic Schr{\"o}\-ding\-er equation 
\begin{equation}
\ii{\dot \psi} = -\frac{1}{2m}  \nabla^2 \psi \,,
\label{s.1}
\end{equation}
which can be derived from the Lagrangian 
\begin{equation}
L =  \int d^3 x \left( \frac{\ii}{2}\left( \psi^*{\dot \psi}  - \psi {\dot \psi}^*\right) + \frac{1}{2m} \psi^* \nabla^2 \psi \right)\,.
\label{s.2}
\end{equation}

In order to pass to the Hamiltonian formulation it is convenient to express $\psi$ in terms of real and imaginary parts.{\footnote{See \cite{gergely02} for an overview of possible derivations of the Hamiltonian formulation for the Schr\"odinger field.}} We write $\psi=\psi_r + \ii \psi_i$. The canonically conjugate momenta, the Hamiltonian and the constraints are respectively given by  
\begin{equation}
\Pi_{\psi_r} = \frac{\delta L }{\delta {\dot \psi_r} } = \psi_i \,, \quad \Pi_{\psi_i} = \frac{\delta L }{\delta {\dot \psi_i} } = -\psi_r\,,
\label{s.3}
\end{equation}
\begin{equation}
H=-\frac{1}{2m} \int d^3 x \left(\psi_r \nabla^2 \psi_r + \psi_i \nabla^2 \psi_i \right)\,,
\label{s.4}
\end{equation}
\begin{equation}
\chi_1 = \Pi_{\psi_r} - \psi_i \,, \quad \chi_2 = \Pi_{\psi_i} + \psi_r \,.
\label{s.5}
\end{equation}
Since $[\chi_1({\bf x}),\chi_2({\bf y})]_P = -2\delta({\bf x} - {\bf y})$, the constraints are second class.

As explained in section \ref{canonicalquantization}, we proceed by separating the unconstrained variables from the constraints by performing a canonical transformation. Such a canonical transformation is given by
\begin{align}
&\psi_r = \frac{1}{\sqrt{2}} \left(\phi + \phi' \right)\,,\qquad \Pi_{\psi_r} = \frac{1}{\sqrt{2}} \left(\Pi_{\phi} + \Pi_{\phi'} \right) \,, \nonumber\\
&\psi_i = \frac{1}{\sqrt{2}} \left(\Pi_{\phi} - \Pi_{\phi'} \right) \,,\qquad \Pi_{\psi_i} = \frac{1}{\sqrt{2}} \left(\phi' - \phi \right)  \,.
\label{s.6}
\end{align}
In terms of the new canonical variables $\phi,\Pi_{\phi},\phi'$ and $\Pi_{\phi'}$, the constraints read $\phi'=\Pi_{\phi'}=0$, so that $\phi$ and $\Pi_{\phi}$ are unconstrained variables. The Hamiltonian for the unconstrained variables reads 
\begin{equation}
H=-\frac{1}{4m} \int d^3 x \left( \Pi_{\phi} \nabla^2 \Pi_{\phi} + \phi \nabla^2\phi  \right)\,.
\label{s.7}
\end{equation} 
The corresponding Hamilton equations reduce to the second order differential equation
\begin{equation}
{\ddot \phi} +  \frac{1}{4m^2}\nabla^4 \phi =0\,. 
\label{s.7.01}
\end{equation}
Note that this equation of motion corresponds to the square of the non-relativistic dispersion relation.

Quantization proceeds by associating field operators $\widehat{\phi}$ and $\widehat{\Pi}_{\phi}$ to the canonical variables and by imposing the standard commutation relations
\begin{equation}
[\widehat{\phi}({\bf x}), \widehat{\Pi}_{\phi}({\bf y})]=\ii\delta({\bf x}-{\bf y})\,.
\label{s.7.1}
\end{equation}
The corresponding Hamiltonian operator is given by
\begin{equation}
{\widehat H}=-\frac{1}{4m} \int d^3 x \left( \widehat{\Pi}_{\phi} \nabla^2 \widehat{\Pi}_{\phi}  + \widehat{\phi} \nabla^2\widehat{\phi}  \right)\,.
\label{s.8}
\end{equation}
Using the functional Schr{\"o}\-ding\-er representation
\begin{equation}
\widehat{\phi} \to \phi \,, \quad \widehat{\Pi}_{\phi} \to -\ii \frac{\delta }{\delta \phi}\,,
\label{s.12}
\end{equation}
the functional Schr{\"o}\-ding\-er equation
\begin{equation}
\ii\frac{\pa \Psi}{\pa t} = \int d^3 x \int d^3 y  \left(-\frac{1}{2} \frac{\delta }{\delta \phi({\bf x})} h({\bf x},{\bf y})\frac{\delta }{\delta \phi({\bf y})}  + \frac{1}{2} \phi({\bf x})h({\bf x},{\bf y})\phi({\bf y})  \right)\Psi 
\label{s.13}
\end{equation}
is obtained, where 
\begin{equation}
h({\bf x},{\bf y}) = -\frac{1}{2m}\nabla^2 \delta({\bf x}-{\bf y})\,.
\label{s.14}
\end{equation}
A field beable $\phi$ can now be introduced, which evolves according to 
\begin{equation}
{\dot \phi}({\bf x}) = \int d^3 y h({\bf x},{\bf y}) \frac{\delta S}{\delta \phi({\bf y})} =  -\frac{1}{2m}\nabla^2 \frac{\delta S}{\delta \phi({\bf x})} 
\label{s.15}
\end{equation}
and for which the quantum equilibrium density is given by $|\Psi(\phi,t)|^2$.

\subsection{Comparison with alternative formulations}
\subsubsection{Alternative Hamiltonian formulation}
Usually the quantized Schr{\"o}\-ding\-er field is formulated in terms of the field operators ${\widehat \psi}$ and ${\widehat \psi^*}$, which satisfy the commutation relations
\begin{equation}
[{\widehat \psi}({\bf x}),{\widehat \psi^*}({\bf y})] = \delta({\bf x}-{\bf y}) \,.
\label{s.16}
\end{equation}
A systematic way of deriving these commutation relations involves constructing the Dirac bracket for the classical fields $\psi$ and $\psi^*$ (which are treated as independent fields), and their corresponding momenta $\Pi_{\psi}$ and $\Pi_{\psi^*}$  \cite{gergely02}, which are given by 
\begin{equation}
\Pi_{\psi}=\ii\psi^*/2\,, \qquad \Pi_{\psi^*}=-\ii\psi/2\,.
\label{s.15.1}
\end{equation}

The formulation in terms of the operators  ${\widehat \psi}$ and ${\widehat \psi^*}$ is equivalent to the formulation in terms of $\widehat{\phi}$ and $\widehat{\Pi}_{\phi}$. The equivalence is obtained by using the relations ${\widehat \psi}={\widehat \psi}_r +\ii {\widehat \psi}_i$ and ${\widehat \psi^*}={\widehat \psi}_r - \ii {\widehat \psi}_i$, by using the operator form of the canonical transformation (\ref{s.6}), i.e.\ ${\widehat \psi}_r=(\widehat{\phi} + \widehat{\phi}')/ \sqrt{2}$ and ${\widehat \psi}_i =(\widehat{\Pi}_{\phi} - \widehat{\Pi}_{\phi'}) / \sqrt{2}$, and by imposing the constraints as operator identities, i.e.\ $\widehat{\phi}' =\widehat{\Pi}_{\phi'}=0$. One can easily see that these relations imply the equivalence of the commutation relations (\ref{s.7.1}) and (\ref{s.16}).

In terms of the field operators ${\widehat \psi}$ and ${\widehat \psi^*}$ the Hamiltonian operator (\ref{s.8}) has the familiar form
\begin{equation}
{\widehat H} =  -\frac{1}{2m} \int d^3 x {\widehat \psi^*}\nabla^2 {\widehat \psi} + E_0 \,,
\label{s.17}
\end{equation}
where
\begin{equation}
E_0= \frac{1}{4m} \int d^3 x \int d^3 y {\boldsymbol \nabla}_x \delta({\bf x}-{\bf y}) \cdot {\boldsymbol \nabla}_x \delta({\bf x}-{\bf y}) 
\label{s.18}
\end{equation}
is an infinite vacuum energy. It is common practice to drop the infinite constant. This is usually done by assuming a suitable operator ordering. Here we will leave the operator ordering as it is. The removal of the infinite constant in the Hamiltonian operator for ${\widehat \psi}$ and ${\widehat \psi^*}$ would imply the addition of this infinite constant in the Hamiltonian operator for $\widehat{\phi}$ and $\widehat{\Pi}_{\phi}$. The appearance of this divergent term clearly shows the need for a regularization. 

So our formulation of the quantized Schr{\"o}\-ding\-er field is equivalent to the more familiar formulation in terms of the field operators ${\widehat \psi}$ and ${\widehat \psi^*}$. In retrospect, we could have developed the pilot-wave model starting from this formulation, by realizing the commutation relations of the operators ${\widehat \psi}$ and ${\widehat \psi^*}$ by the representation
\begin{equation}
{\widehat \psi} \to \frac{1}{\sqrt{2}}\left( \phi + \frac{\delta }{\delta \phi} \right)\,,\quad  {\widehat \psi^*} \to \frac{1}{\sqrt{2}}\left( \phi - \frac{\delta }{\delta \phi} \right) \,.
\label{s.18.1}
\end{equation}

\subsubsection{Alternative pilot-wave approaches}\label{alternativepilotwaveapproaches}
Note that by introducing a field beable $\phi({\bf x})$ for the field operator $\widehat{\phi}({\bf x})$ we have in fact introduced a beable for the operator ${\widehat \psi}_r({\bf x})$, namely the field $\psi_r=\phi/\sqrt{2}$. Alternatively, we could have introduced a field beable $\Pi_\phi({\bf x})$ for the momentum field operator $\widehat{\Pi}_{\phi}({\bf x})$, which is related to the operator ${\widehat \psi}_i({\bf x})$. This underdetermination can be traced back to the underdetermination in the canonical transformation that separates the unconstrained degrees of freedom from the constraints, cf.\ section \ref{canonicalquantization}, since one can always perform a canonical transformation that interchanges the role of fields and momenta. Similar underdeterminations were pointed out for the case of the electromagnetic field \cite{baumann86}. 

There is also a way to introduce beables for ${\widehat \psi}_r$ and ${\widehat \psi}_i$ simultaneously (instead of for either one of the two), and hence for ${\widehat \psi}$. For example, starting from the pilot-wave model presented above, with beable $\psi_r=\phi/\sqrt{2}$ corresponding to ${\widehat \psi}_r$, one could also introduce a beable $\psi_i$ corresponding to ${\widehat \psi}_i$, by means of Holland's local expectation value, cf.\ section \ref{construction of a pilot-wave model}. In the field representation \eqref{s.18.1} the operator ${\widehat \psi}_i$ is represented by the operator $\frac{-\ii}{\sqrt{2}} \frac{\delta}{\delta \phi}$, so that the corresponding beable is given by
\begin{equation}
\psi_i({\bf x},t) = {\textrm{Re}} \frac{\Psi^*(\phi,t) \left( \frac{-\ii}{\sqrt{2}} \frac{\delta}{\delta \phi({\bf x})} \right) \Psi(\phi,t)}{|\Psi(\phi,t)|^2} \Bigg|_{\phi=\phi({\bf x},t)} = \frac{1}{\sqrt{2}} \frac{\delta S(\phi,t)}{\delta \phi({\bf x})}\Bigg|_{\phi=\phi({\bf x},t)}\,,
\label{s.18.1.1}
\end{equation}
where the expressions on the right hand side are evaluated for the actual value of the field beable $\phi({\bf x},t)$. The beable $\psi$ for the operator $\widehat \psi$ is then given by 
\begin{equation}
\psi = \psi_r + \ii \psi_i =  \frac{1}{\sqrt{2}} \left(\phi + \ii \frac{\delta S}{\delta \phi}  \right) \,,
\label{s.18.1.2}
\end{equation}
where the right hand side is evaluated for the actual beable configuration $\phi$.

Another way to obtain a beable $\psi$ for ${\widehat \psi}$ would be to first introduce a beable $\Pi_\phi$ for the momentum field operator $\widehat{\Pi}_{\phi}$ in the standard way (using the continuity equation to obtain a guidance equation), and hence for ${\widehat \psi}_i$, and to subsequently introduce a additional beable for ${\widehat \psi}_r$ by means of Holland's local expectation value. The pilot-wave model as such obtained would be different from any of the models discussed above.

In the approach of Takabayasi \cite{takabayasi52} and Holland \cite{holland881} and \cite[pp.\ 449-451]{holland93b}, the beable is also given by a complex field $\psi({\bf x})$, with a form similar to that of \eqref{s.18.1.2}. The difference merely arises from the fact that Takabayasi and Holland use a representation of the field operators that is different from \eqref{s.18.1}.

All of the possible pilot-wave approaches just described concern a field in physical space, so that they provide a clear ontology and hence there is no a priori reason to prefer one ontology above the other. Of course, it could be that either ontologies fails to yield an image of macroscopic objects and hence fails to reproduce the standard quantum predictions and could be excluded on this basis. Note also that theories with different ontologies that reproduce the standard quantum predictions are evidently empirically indistinguishable. In quantum non-equilibrium these theories do not yield the standard quantum predictions and they might under these circumstances be distinguishable. 

However, the Schr{\"o}\-ding\-er field describes identical bosonic particles,{\footnote{Note that in writing about particles here, we have employed the ordinary parlance of standard quantum theory; we are not referring to actual point-particles. Throughout the paper, it should always be clear from the context what meaning we have in mind.}} so that, despite the many applications of the theory, it can not really faithfully describe macroscopic objects, like pointer needles; since such objects are composed out of fermions, they should be described by a fermionic field theory. Therefore we can not really discuss how our pilot-wave model in terms of the field beable is able to reproduce the quantum predictions. We return to this in section \ref{someelementarystates}.

\subsection{Valentini's suggestion for dealing with constraints}\label{valentinisway}
\subsubsection{Dirac field}\label{valentiniswaydiracfield}
In \cite{valentini92,valentini96} Valentini tried to develop a pilot-wave model for the quantized Dirac field (which describes relativistic spin-1/2 particles). With the quantization of the Dirac field constraints appear that are similar to the ones we encountered in the case of the Schr{\"o}\-ding\-er field. It is interesting to consider how Valentini dealt with the obstacle of constraints, because it is very similar in spirit to the approach adopted here. In fact, since Valentini was dealing with fermionic quantization, he was facing additional obstacles in the construction of a pilot-wave model. However, these additional obstacles will be discussed in detail in section \ref{valentinismodelfermions}. 

For the Dirac field the constraints read 
\begin{equation}
\Pi_{\psi_a}=\ii\psi^*_a/2\,, \qquad \Pi_{\psi^*_a}=-\ii\psi_a/2\,.
\label{s.18.2}
\end{equation}
Note that these constraints are completely similar to the constraints (\ref{s.15.1}) for the Schr\"o\-din\-ger field; there is just an extra spinor index in this case. Valentini realized that it would make no sense to impose the guidance equations
\begin{equation}
\frac{\ii\psi^*_a}{2} = \frac{\delta S}{\delta \psi_a}\,,
\label{s.18.3}
\end{equation}
since rather than telling us the rate of change of the field beable, the functional derivative $\delta S/\delta \psi_a$ would tell us the actual value of the field beable. Valentini reasoned that the appearance of the constraints, and hence the reason for the difficulties, was due to the fact that the Dirac equation is first order in time derivatives. As a solution, Valentini proposed to use a second order formulation of the theory instead. The second order equivalent of the first order Dirac theory is the Van der Waerden theory. Since the Hamiltonian formulation of the Van der Waerden theory does not involve constraints, the above problem disappears.

Let us comment on this. First of all, we want to say a word on the particular choice of the guidance equation (\ref{s.18.3}). Valentini obtained the guidance equation by equating $\delta S / \delta \phi_r$ and $ \Pi_r(\phi,\dot{\phi})$, where $S$ is the phase of the wave functional and where $\Pi_r(\phi,\dot{\phi})$ are the classical momenta conjugate to the fields $\phi_r$. This prescription is inspired by the Hamilton-Jacobi formulation of Hamiltonian systems. In a number of familiar situations, like e.g.\ in the non-relativistic pilot-wave theory of \db\ and Bohm, this prescription yields the same guidance equation as the one that can be derived by considering the continuity equation for the probability density $|\Psi|^2$. However, this prescription does not work in general. The Dirac field is just one example where it fails. In Appendix \ref{appendixa}, another simple example is given which, unlike the Dirac theory, involves bosonic quantization and has a Hamiltonian formulation that does not involve constraints. 

Despite of this, Valentini's insight of using a second order formulation of the theory is in agreement with our approach. In our approach the constraints were separated from the unconstrained variables and only the latter were quantized. In effect, the first order Schr{\"o}\-ding\-er equation was thereby replaced by a second order equation \eqref{s.7.01} for the field by $\phi$, which is in accordance with Valentini's suggestion.

\subsubsection{String field theory}
In \cite{weingard95} Weingard tried to develop a pilot-wave model for the second quantized bosonic string (see also \cite{huggett99} for a summary). In doing so, Weingard also faced the problem of constraints. Weingard adopted the line of reasoning of Valentini and suggested to look for a second order reformulation of the theory. Such a second order reformulation was unknown to Weingard, which forced him to halt the attempt to develop a pilot-wave model. However, the theory of the second quantized bosonic string is formally similar to the quantized Schr{\"o}\-ding\-er field and therefore a pilot-wave model can be developed in a similar way as for the Schr{\"o}\-ding\-er field. Let us sketch how this would go. 

First recall the basic ingredients of string theory in the light-cone gauge. A classical string in $(D-1)$-dimensional space traces out a two-dimensional surface in $D$-dimensional space-time, called the world-sheet. The world-sheet can be represented by the coordinate functions $x^{\mu}(\sigma,\tau)$. It can also be represented in terms of light-cone coordinates $x^+,x^-,x^i$, with $i=1,\dots,D-2$, where $x^+$ and $x^-$ are coordinates along the light-cone and $x^i$ the $D-2$ remaining spatial coordinates. In the light-cone gauge the parameter $\tau$ is proportional $x^+$ and the coordinate function $x^-$ can be expressed in terms of the $x^i$ and a constant of the motion $x^-_0$. In the resulting Hamiltonian formulation, the parameter $\tau$ plays the role of time and the dynamical degrees of freedom are the functions $x^i(\sigma)$. 

In the first quantized theory, the states are functionals $\psi(x^i(\sigma),\tau)$ \cite{hatfield91} and a pilot-wave model can be constructed by introducing beables $x^i(\sigma)$ which represent a curve in $D-2$ dimensions \cite{weingard95}. (Hereby the degree of freedom $x^-_0$ has been ignored. We leave aside the question whether or not an additional beable should be introduced for $x^-_0$ or $x^-$.) 

In the second quantized theory, one has operators ${\widehat \psi}(x^i(\sigma)) $ and ${\widehat \psi}^*(x^i(\sigma))$, which satisfy the commutation relations
\begin{equation}
[{\widehat \psi}(x),{\widehat \psi}^*(y )] = \delta(x - y) 
\label{s.18.5}
\end{equation}
and the Hamiltonian is of the form
\begin{equation}
{\widehat H} = \int {\mathcal D} x {\mathcal D} y {\widehat \psi}^*(x) h(x,y) {\widehat \psi}(y)\,,
\label{s.18.6}
\end{equation}
where $h(x,y)$ is some kernel \cite{weingard95}. These equations are formally similar to the ones for the quantized Schr{\"o}\-ding\-er field, cf.\ (\ref{s.16}) and (\ref{s.17}), and hence a pilot-wave model can be developed in a similar way as for the Schr{\"o}\-ding\-er field. Without going into detail, this would result in a beable $\phi(x^i,\tau)$ that is being guided by a wave functional $\Psi(\phi(x^i),\tau)$. The beable $\phi(x^i,\tau)$ would then be a real-valued functional, defined on the space of curves $x^i(\sigma)$ in a $(D-2)$-dimensional space. Whether such a beable is capable of providing an image of macroscopic objects (so that, in particular, measurement results would be recorded and displayed in the beable) is of course an open question.

\subsection{Some elementary states}\label{someelementarystates}
As explained in section \ref{reproducingquantumpredictions}, in order to find out whether a pilot-wave model is capable of reproducing the standard quantum predictions, it is important to consider which fields make up the support of wave functionals representing macroscopic objects. In particular, wave functionals representing macroscopically distinct states should have approximately disjoint supports. However, as mentioned before, the quantized Schr{\"o}\-ding\-er field describes identical bosonic particles and can therefore not really faithfully describe macroscopic objects like pointer needles. Hence we can not really discuss how the pilot-wave model is able to reproduce the quantum predictions. Nevertheless, it is still interesting to look at some familiar states and consider the asociated field distributions. This is because similar results apply to other bosonic field theories discussed in this paper, like for example the electromagnetic field.

We will subsequently consider the vacuum state, coherent states and states which describe a system with a definite number of particles. Since we are merely interested in the associated field distributions we will not give the associated guidance equations. Some discussion of the guidance equations corresponding to these states, in the context of the scalar field and the electromagnetic field, can be found in respectively \cite{bohm87b,bohm93,holland93b,holland93a,lam941,lam942} and \cite{kaloyerou94}.

\subsubsection{Vacuum state}
As is well known the operators ${\widehat \psi^*}({\bf x})$ and ${\widehat \psi}({\bf x})$ respectively create and annihilate a particle at the position ${\bf x}$. The operators ${\widehat a}^\dagger({\bf k})$ and ${\widehat a}({\bf k})$, which are introduced through the relations
\begin{equation}
{\widehat \psi}({\bf x}) = \frac{1}{(2\pi)^{3/2}} \int d^3 k e^{\ii {\bf k} \cdot {\bf x}}{\widehat a}({\bf k}) \,,\quad {\widehat \psi^*}({\bf x}) = \frac{1}{(2\pi)^{3/2}} \int d^3 k e^{-\ii {\bf k} \cdot {\bf x}}{\widehat a}^\dagger({\bf k})\,,
\label{s.19}
\end{equation}
respectively create and annihilate a particle with momentum ${\bf k}$. The vacuum, which contains no particles, is represented by the state $|\Psi_0(t)\ra$. It satisfies ${\widehat \psi}({\bf x})|\Psi_0(t)\ra=0 $ for all ${\bf x}$ or equivalently ${\widehat a}({\bf k}) |\Psi_0(t)\ra=0 $ for all ${\bf k}$. 

In the functional Schr{\"o}\-ding\-er picture, we find by using ({\ref{s.18.1}}) that the vacuum wave functional $\Psi_0(\phi,t)=\la \phi |\Psi_0(t) \ra$ is determined by
\begin{equation}
\left( \phi({\bf x}) + \frac{\delta }{\delta \phi({\bf x})} \right)\Psi_0(\phi,t) = 0\,.
\label{s.20}
\end{equation}
The solution reads
\begin{equation}
\Psi_0(\phi,t)=N \exp  \left(-\frac{1}{2} \int d^3x \phi({\bf x})^2 - iE_0 t \right)\,,
\label{s.21}
\end{equation}
where $N=\prod_{\bf x} \pi^{-1/4}$ is an infinite normalization constant. It can easily be verified that $\Psi_0(\phi,t)$ satisfies the functional Schr{\"o}\-ding\-er equation. The ground state is a Gaussian wave functional which is centered around the field configuration $\phi=0$. 

\subsubsection{Coherent state}
Another state of interest is the coherent state $|\Psi_\alpha(t) \ra$ \cite{glauber63}. The coherent state $|\Psi_\alpha(t)\ra$ satisfies 
\begin{equation}
{\widehat a}({\bf k}) |\Psi_\alpha (t)\ra = \alpha({\bf k})e^{-\ii k^2t/2m}|\Psi_\alpha(t) \ra.
\label{s.23}
\end{equation}
In case we can take the Fourier transform of $\alpha({\bf k})$, we also have 
\begin{equation}
{\widehat \psi} ({\bf x}) |\Psi_\alpha (t)\ra = \alpha({\bf x},t) |\Psi_\alpha (t)\ra\,,
\label{s.24}
\end{equation}
where
\begin{equation}
\alpha({\bf x},t) = \frac{1}{(2\pi)^{3/2}}  \int d^3 k  e^{\ii {\bf k} \cdot {\bf x}-\ii k^2t/2m} \alpha({\bf k})\,.
\label{s.25}
\end{equation}

In the functional Schr{\"o}\-ding\-er picture, the coherent state $\Psi_\alpha(\phi,t)=\la \phi |\Psi_\alpha(t)\ra$ satisfies
\begin{equation}
\frac{1}{\sqrt{2}}\left( \phi({\bf x}) + \frac{\delta }{\delta \phi({\bf x})} \right) \Psi_\alpha(\phi,t) = \alpha({\bf x},t) \Psi_\alpha(\phi,t) \,.
\label{s.26}
\end{equation}
The solution is given by
\begin{equation}
\Psi_\alpha(\phi,t) =N \exp \left( -\frac{1}{2} \int d^3 x \left(\phi({\bf x}) -\sqrt{2} \alpha({\bf x},t)  \right)^2 -\ii E_0t\right)\,,
\label{s.27}
\end{equation}
where $N=(\prod_{\bf x} \pi^{-1/4})\exp(- \int  d^3 x \alpha_i^2)$ is a normalization constant and where $\alpha = \alpha_r + \ii \alpha_i$ was used, with $\alpha_r$ and $\alpha_i$ real.

Writing 
\begin{multline}
\Psi_\alpha(\phi,t) = \left(\prod_{\bf x}  \pi^{-1/4} \right)  \exp \bigg(  -\frac{1}{2} \int d^3 x \left(\phi({\bf x}) -\sqrt{2} \alpha_r({\bf x},t)  \right)^2    \\
 + \ii \sqrt{2}  \int d^3 x \left(\phi({\bf x}) - \sqrt{2} \alpha_r({\bf x},t) \right)\alpha_i({\bf x},t)    -\ii E_0t \bigg)\,,
\label{s.28}
\end{multline}
we see that the field density $|\Psi_\alpha(\phi,t)|^2$ is a Gaussian centered around the field $\sqrt{2} \alpha_r({\bf x},t)$. In the case the function $\alpha$ is zero we have the ground state wave functional. 

It is clear that it is sufficient that the real parts of $\alpha_1$ and $\alpha_2$ differ significantly in just a region of physical space in order for the wave functionals $\Psi_{\alpha_1}$ and $\Psi_{\alpha_2}$ to be non-overlapping. Nevertheless, these wave functionals do not describe macroscopic systems in non-relativistic quantum theory. Of course we are considering a theory which describes only identical bosonic particles, but there is an additional reason. The reason is that we think of a macroscopic system, like a pointer needle, as a system composed out of an approximately definite number of particles. To the contrary, the coherent state is a superposition of states with all possible particle numbers. On the other hand, if we consider electromagnetism, then coherent states for which the mean photon number is sufficiently large will describe familiar macroscopic states of light. Therefore the above discussion becomes important for the pilot-wave theory for the electromagnetic field (see section \ref{freeemfield}). States for the Schr{\"o}\-ding\-er field describing a definite particle number are discussed in the following section.

\subsubsection{State describing a definite number of particles}
An arbitrary state describing $n$ particles is given by
\begin{equation}
|\Psi_{\psi}(t) \ra =\frac{1}{\sqrt{n!}}  \int d^3 x_1 \dots \int d^3 x_n \psi({\bf x}_1,\dots,{\bf x}_n,t ) {\widehat \psi^*} ({\bf x}_1) \dots {\widehat \psi^*} ({\bf x}_n) |\Psi_0(t)\ra\,,
\label{s.29}
\end{equation}
where the completely symmetric expansion coefficients $\psi({\bf x}_1,\dots,{\bf x}_n,t )$ are given by  
\begin{equation}
\psi({\bf x}_1,\dots,{\bf x}_n,t ) = \frac{1}{\sqrt{n!}}  \la \Psi_0(t) |  {\widehat \psi} ({\bf x}_n) \dots {\widehat \psi} ({\bf x}_1) |\Psi_{\psi}(t) \ra \,.
\label{s.29.1}
\end{equation}
The requirement that $|\Psi_{\psi}(t) \ra$ satisfies the field theoretical Schr{\"o}\-ding\-er equation implies that $\psi({\bf x}_1,\dots,{\bf x}_n,t )$ satisfies the non-relativistic Schr{\"o}\-ding\-er equation. Consider the wave function $\psi({\bf x}_1,\dots,{\bf x}_n,t )$ to be normalized to one, so that the state $|\Psi_{\psi}(t) \ra$ is normalized to one.

For a single particle the wave functional reads
\begin{eqnarray}
\Psi_{\psi} (\phi,t)&=& \int d^3x   \psi({\bf x},t)\frac{1}{\sqrt{2}}\left( \phi({\bf x}) - \frac{\delta }{\delta \phi({\bf x})} \right) \Psi_0(\phi,t) \nonumber\\
&=& \sqrt{2}\int d^3x   \psi({\bf x},t) \phi({\bf x})\Psi_0(\phi,t)  \,.
\label{s.30}
\end{eqnarray}
The wave functional corresponding to any number of particles can be written in a similar fashion.

In the pilot-wave theory of \db\ and Bohm for non-relativistic quantum theory, two states representing a single particle localized at different regions in physical space are non-overlapping. With the field ontology this does not seem true anymore. There is a strong indication that wave functionals describing a single particle will in general have significant overlap, regardless of whether these states describe localized particles.{\footnote{Valentini, private communication.}} To see this consider the following.

The field density corresponding to a one-particle state $\Psi_{\psi}$ is given by
\begin{equation}
\left| \Psi_{\psi} (\phi) \right|^2 = 2 |\alpha|^2  |\Psi_0(\phi) |^2 \,,
\label{s.31}
\end{equation}
where
\begin{equation}
\alpha = \int d^3x \psi({\bf x}) \phi({\bf x}) \,.
\label{s.32}
\end{equation}
The density reaches a maximum for fields that satisfy $\de |\Psi_{\psi}|^2 / \de \phi = 0$. This implies that either{\footnote{A similar calculation was done in \cite{valentini92} for the scalar field.}} 
\begin{equation}
\alpha = 0 
\label{s.32.1}
\end{equation}
or
\begin{equation}
\phi({\bf x}) = \frac{1}{2}\left( \frac{\psi({\bf x})}{\alpha}  + \frac{\psi^*({\bf x})}{\alpha^*} \right)\,.
\label{s.33}
\end{equation}
For a field $\phi$ for which $\alpha = 0$, the density $\left| \Psi_{\psi} (\phi) \right|^2$ is zero and hence for such a field a minimum is reached. Maxima are reached for fields which satisfy \eqref{s.33}, i.e.\ for fields which are a linear combination of $\psi$ and $\psi^*$.

Consider now the superposition $\Psi_{N(\psi_1 + \psi_2)}=N\left( \Psi_{\psi_1} + \Psi_{\psi_2}\right)$, of two one-particle states $\Psi_{\psi_1}$ and $\Psi_{\psi_2}$, with $N$ a normalization factor. This state has maxima for fields that are linear combination of $\psi_1 + \psi_2$ and $\psi^*_1 + \psi^*_2$. On the other hand, if we assume that $\Psi_{\psi_1}$ and $\Psi_{\psi_2}$ have negligible overlap, then
\begin{equation}
|\Psi_{N(\psi_1 + \psi_2)}|^2 \approx N^2 \left( |\Psi_{\psi_1}|^2 + |\Psi_{\psi_2}|^2 \right) \,.
\label{s.34}
\end{equation}
The right hand side reaches maxima for fields which are a linear combination of $\psi_1$ and $\psi^*_1$ and for fields which are a linear combination of $\psi_2$ and $\psi^*_2$. These do not correspond to the maxima obtained for $|\Psi_{N(\psi_1 + \psi_2)}|^2$. Because of \eqref {s.34}, this suggests that in general the states $\Psi_{\psi_1}$ and $\Psi_{\psi_2}$ will have significant overlap. 

For clarity, in the special case the wave functionals $\Psi_{\psi_1}$ and $\Psi_{\psi_2}$ represent a particle localized in different regions in physical space, $\psi_1$ and $\psi_2$ will have negligible overlap in physical space. However, as the above analysis shows, this does not imply that the wave functionals $\Psi_{\psi_1}$ and $\Psi_{\psi_2}$ have negligible overlap in field space.

We think of a macroscopic system, like a measurement needle, not only as an approximately localized system, but also as a system composed of a large, but approximately fixed number of particles. Therefore it would actually be more interesting to consider wave functionals for such systems and see in which cases they might be non-overlapping. Recently some numerical simulations have been performed by Schmelzer which indicate a decrease in overlap that goes exponentially with the number of particles \cite{schmelzer10}.


\section{Free electromagnetic field}\label{freeemfield}
We now turn to the free electromagnetic field. In this and the following sections, the components of four-vectors are labeled by Greek indices, e.g.\  $\mu$ and $\nu$, and the components of spatial three-vectors are labeled by Latin indices, e.g.\  $i$ and $j$. For these indices we use Einstein's summation convention.

We consider two pilot-wave approaches to the electromagnetic field. One is by Bohm and takes a tranverse potential as beable. The other one is by Valentini and takes an equivalence class of potentials as beable. While these ontologies are different, these models appear to be empirically equivalent, not only at the level of quantum equilibrium, but also for non-equilibrium distributions.

\subsection{Hamiltonian formulation}
The Lagrangian and the Euler-Lagrange equations of motion for the free classical electromagnetic field, described by the vector potential $A^{\mu}=(A_0,A_i)$, are respectively given by
\begin{equation}
L =  -\frac{1}{4}\int d^3 x  F^{\mu \nu}F_{\mu \nu}\,,\quad F^{\mu \nu} = \partial^\mu A^{\nu} - \partial^\nu A^{\mu}\,,
\label{e.1}
\end{equation}
\begin{equation}
\partial_\mu F^{\mu \nu} = 0\,.
\label{e.2}
\end{equation}
The theory has a local symmetry 
\begin{equation}
A^{\mu} \to A^{\mu} - \partial^\mu  \theta\,.
\label{e.3}
\end{equation}
The arbitrariness of $\theta$ as a function of space-time signals gauge freedom.{\footnote{It is straightforward to show that the transformations \eqref{e.3} are all gauge transformations in the sense defined in section \ref{canonicalquantization}. More details can be found in \cite{struyve09c}.}} This gives rise to first class constraints in the Hamiltonian picture.

Indeed, in the Hamiltonian picture we have \cite{hanson76,sundermeyer82,gitman90,henneaux91}:
\begin{equation}
\Pi_{A_0} = \frac{\delta L }{\delta {\dot A_0} } = 0\,, \quad \Pi_{A_i} = \frac{\delta L }{\delta {\dot A_i} } = {\dot A}_i +\partial_i A_0 \,,
\label{e.4}
\end{equation}
\begin{equation}
H = \int d^3 x \left( \frac{1}{2}\Pi_{A_i}\Pi_{A_i} + \frac{1}{4} F_{ij}F_{ij} +A_0\partial_i \Pi_{A_i} \right)\,,\quad  F_{ij}=\partial_i A_j - \partial_j A_i\,,
\label{e.5}
\end{equation}
\begin{equation}
\chi_{1} = \Pi_{A_0} \,, \quad \chi_{2} = \partial_i \Pi_{A_i}\,,
\label{e.5.1}
\end{equation}
where the constraints $\chi_{1}$ and $\chi_{2}$ are first class. Since $\Pi_{A_i}={\dot A}_i +\partial_i A_0 = -E_i$, with $E_i$ the electric field, the second constraint can be written as $\pa_i E_i =0$, which is just one of the free Maxwell equations, namely the Gauss law.

\subsection{Bohm's approach: transverse potential as beable}
In presenting his pilot-wave model for the electromagnetic field, Bohm started with imposing the Coulomb gauge $\partial_i A_i=0$ \cite{bohm52b}. (From the classical equations of motion it then follows that $\nabla^2 A_0 = 0$, so that, assuming that $A_0$ vanishes sufficiently fast at spatial infinity, $A_0=0$.) Bohm then expressed the transverse parts of the fields in terms of Fourier modes, which form unconstrained canonical pairs. The pilot-wave model was then introduced for the quantum field theory that results from quantizing those Fourier modes. Belinfante \cite[pp.\ 198-209]{belinfante73} and Kaloyerou \cite{kaloyerou85,kaloyerou94,kaloyerou96}, who discuss Bohm's model in detail, adopted a similar approach. 

The quantum field theory for the transverse modes of the field can actually be obtained more directly by identifying unconstrained canonical pairs as gauge independent degrees of freedom and by quantizing these (as discussed in section \ref{canonicalquantization}).{\footnote{In view of the last paragraph of section \ref{canonicalquantization}, it is not surprising that the same theory can be obtained without gauge fixing.}} We follow this approach to present Bohm's model because it immediately makes clear that the model is gauge independent, not only at the statistical level in quantum equilibrium, but also at the fundamental level. This also immediately answers the questions raised by Kaloyerou \cite{kaloyerou94,kaloyerou96} concerning the gauge invariance of the pilot-wave model.

\subsubsection{Quantization of gauge independent degrees of freedom}
In order to quantize only the gauge independent degrees of freedom, the first step is to perform a canonical transformation so that in terms of the new canonical variables, the constraints are given by some of the momenta. Since the momentum $\Pi_{A_0}$ is constrained to be zero, $A_0$ can immediately be identified as a gauge variable. In order to identify the remaining gauge variable, we write $A_i = A^T_i + A^L_i$, where $A^T_i$ and $A^L_i$ are respectively the transverse and longitudinal part of $A_i$, defined by 
\begin{equation}
A^T_i =  \left(\delta_{ij} -  \frac{\partial_i \partial_j }{\nabla^2}\right) A_j\,, \quad A^L_i =  \frac{\partial_i \partial_j }{\nabla^2} A_j\,,
\label{e.7}
\end{equation}
and where the integral operator $\nabla^{-2}$ is given by
\begin{equation}
\frac{1}{\nabla^2} f({\bf x}) = -\int d^3 y \frac{f({\bf y})}{4\pi|{\bf x}-{\bf y}|}\,.
\label{e.8}
\end{equation}
The fields $A^T_i$ and $\Pi^T_{A_i}$ have vanishing Poisson brackets with the constraints and hence they are gauge independent degrees of freedom (as is well known \cite{dirac67,hanson76,sundermeyer82,gitman90,henneaux91}). The field $A^L_i$ does not have vanishing Poisson brackets with the constraint $\chi_2$ and is the remaining gauge degree of freedom. 

In order to write the constraint $\chi_2=\partial_i \Pi_{A_i}$ and the gauge degree of freedom $A^L_i$ in terms of a canonical pair, the following canonical transformation of the variables $A_i$ and $\Pi_{A_i}$ is performed: 
\begin{align}
A_i({\bf x}) &= \frac{1}{(2\pi)^{3/2}} \sum^3_{l=1} \int d^3 k e^{\ii {\bf k} \cdot {\bf x}} \varepsilon^l_i({\bf k}) q_l({\bf k})\,, \nonumber\\  
\Pi_{A_i}({\bf x})  &=  \frac{1}{(2\pi)^{3/2}} \sum^3_{l=1} \int d^3 k e^{-\ii {\bf k} \cdot {\bf x}} \varepsilon^l_i({\bf k}) \Pi_{q_l}({\bf k})\,, 
\label{e.9}
\end{align}
where the new (complex) variables $q_l({\bf k})$ and $\Pi_{q_l}({\bf k})$ form canonical pairs. We also introduced the (real) orthonormal polarization vectors $\varepsilon^l_i({\bf k})$, $l=1,2,3$, which we choose to obey $\varepsilon^3_i({\bf k})= k_i/k$ and ${\boldsymbol \varepsilon}^l({\bf k}) = {\boldsymbol \varepsilon}^l(-{\bf k})$ for $l=1,2$. From the last relation and the fact that $A_i$ and $\Pi_{A_i}$ are real, it follows that $q_l({\bf k}) = q^*_l(-{\bf k})$ and $\Pi_{q_l}({\bf k}) = \Pi_{q^*_l}(-{\bf k}) = \Pi^*_{q_l}(-{\bf k})$ for $l=1,2$, with similar relations for $q_3$ and $\Pi_{q_3}$. 

From the relations
\begin{align}
A^T_i &= \frac{1}{(2\pi)^{3/2}} \sum^2_{l=1} \int d^3 k e^{\ii {\bf k} \cdot {\bf x}} \varepsilon^l_i({\bf k}) q_l({\bf k}) \,, & A^L_i &=  \frac{1}{(2\pi)^{3/2}} \int d^3 k e^{\ii {\bf k}  \cdot {\bf x}} \varepsilon^3_i({\bf k}) q_3({\bf k}) \,, \nonumber\\
\Pi^T_{A_i} &= \frac{1}{(2\pi)^{3/2}} \sum^2_{l=1} \int d^3 k e^{-\ii {\bf k} \cdot {\bf x}} \varepsilon^l_i({\bf k}) \Pi_{q_l}({\bf k}) \,, & \Pi^L_{A_i} &=  \frac{1}{(2\pi)^{3/2}} \int d^3 k e^{-\ii {\bf k}  \cdot {\bf x}}\varepsilon^3_i({\bf k}) \Pi_{q_3}({\bf k}) \,,
\label{e.12}
\end{align}
we see that the transverse parts of the fields are described by the variables $q_1,q_2,\Pi_{q_1},\Pi_{q_2}$. The longitudinal parts of the fields are given by $q_3$ and $\Pi_{q_3}$. Because the constraint $ \partial_i \Pi_{A_i}=0$ implies $\Pi^L_{A_i}=0$, it reduces to $\Pi_{q_3} =0$ in terms of the new variables, so that $q_3$ is the remaining gauge variable. 

The Hamiltonian for the gauge independent variables $q_1,q_2,\Pi_{q_1}$ and $\Pi_{q_2}$ is given by
\begin{equation}
H =  \frac{1}{2} \sum^2_{l=1} \int d^3 k \left(\Pi_{q^*_l}\Pi_{q_l} +  k^2 q^*_l q_l \right) \,.
\label{e.12.1}
\end{equation}
As is well known, this resulting theory is readily obtained by imposing the Coulomb gauge and by expressing the transverse fields in terms of Fourier modes. 

After quantization, we obtain the functional Schr{\"o}\-ding\-er equation
\begin{equation}
\ii\frac{\partial \Psi(q_1,q_2,t)}{\partial t} = \frac{1}{2} \sum^2_{l=1} \int d^3 k \left( -\frac{\delta^2}{ \delta q^*_l({\bf k}) \delta q_l({\bf k})} + k^2  q^*_l({\bf k}) q_l({\bf k}) \right)\Psi(q_1,q_2,t)
\label{e.15}
\end{equation}
for the wave\-func\-tion\-al $\Psi(q_1,q_2,t)$.

\subsubsection{Pilot-wave model}\label{bohmspilotwavemodel}
Beables can now be introduced in the way outlined in section \ref{introducingfieldbeables}. They form a pair of fields $(q_1,q_2)$ that evolve according to the guidance equations
\begin{equation}
{\dot q}_l({\bf k})   =  \frac{\delta S}{\delta  q^*_l({\bf k})  }\,, \qquad l=1,2 \,.
\label{e.16}
\end{equation}
The quantum equilibrium density for the field beables is given by $|\Psi(q_1,q_2,t)|^2$. 

The associated ontology in space is given by a transverse vector field $A^T_i({\bf x})$ that is obtained from the fields $(q_1({\bf k}),q_2({\bf k}))$ using the relations (\ref{e.12}). The dynamics can be expressed directly in terms of the transverse field $A^T_i({\bf x})$ as follows. The functional derivative 
\begin{equation}
\frac{\de}{\de A_i({\bf x})} = \frac{1}{(2\pi)^{3/2}} \sum^3_{l=1} \int d^3 k e^{-\ii {\bf k} \cdot {\bf x}} \varepsilon^l_i({\bf k}) \frac{\de}{\de q_l({\bf k})}
\label{e.17}
\end{equation}
can be decomposed into a transverse and a longitudinal part
\begin{align}
\frac{\delta}{\delta A^T_i({\bf x})} &= \left(\delta_{ij} -  \frac{\partial_i \partial_j }{\nabla^2}\right) \frac{\delta}{\delta A_j({\bf x})} = \frac{1}{(2\pi)^{3/2}} \sum^2_{l=1} \int d^3 k e^{- \ii {\bf k} \cdot {\bf x}} \varepsilon^l_i({\bf k}) \frac{\delta}{\delta q_l({\bf k})} \,,\nonumber\\
\frac{\delta}{\delta A^L_i({\bf x})} &=  \frac{\partial_i \partial_j }{\nabla^2} \frac{\delta}{\delta A_j({\bf x})} =\frac{1}{(2\pi)^{3/2}}  \int d^3 k e^{-\ii {\bf k}  \cdot {\bf x}}\varepsilon^3_i({\bf k}) \frac{\delta}{\delta q_3({\bf k})}\,.
\label{e.18}
\end{align}
By inverting the first relation the functional derivatives $\delta/\delta q_l({\bf k})$, $l=1,2$, can be written in terms of $\delta/\delta A^T_i({\bf x})$. Expressing also the fields $q_l({\bf k})$, $l=1,2$,  in terms of the transverse vector field $A^T_i({\bf x})$, by inverting the relations (\ref{e.12}), the functional Schr{\"o}\-ding\-er equation ({\ref{e.15}}) can be rewritten as an equation for the wave functional $\Psi(A^T_i,t)$ 
\begin{equation}
\ii\frac{\partial \Psi}{\partial t} = \frac{1}{2}  \int d^3 x \left( -\frac{\delta^2}{ \delta A^T_i \delta A^T_i} - A^T_i \nabla^2 A^T_i  \right)\Psi \,.
\label{e.20}
\end{equation}
The guidance equation ({\ref{e.16}}) then reads
\begin{equation}
{\dot A}^T_i({\bf x}) = \frac{\delta S}{\delta  A^T_i({\bf x})  }
\label{e.21}
\end{equation}
and the quantum equilibrium distribution $|\Psi(A^T_i,t)|^2 {\mathcal D} A^T$ (where ${\mathcal D} A^T \sim {\mathcal D}q_1{\mathcal D}q_2$).

The ontology in Bohm's pilot-wave model is given by the field $A^T_i$. Just as in the case of the Schr\"odinger field, cf.\ section \ref{alternativepilotwaveapproaches}, and as noted by Baumann \cite{baumann86}, alternative choices are possible. One could for example consider an ontology that is closer to that of classical electromagnetism in terms of the electric and magnetic field. For example one could use Holland's local expectation values to introduce a beable $B_i=\epsilon_{ijk}\partial_jA^T_k$ for the magnetic field operator ${\widehat B}_i$ and a beable $E_i = - \delta S / \delta A^T_i$ for the electric field operator ${\widehat E}_i$.

\subsubsection{Macroscopic states}
Wave functionals representing macroscopically distinct states of the electromagnetic field describe distinct classical electric and magnetic fields. This implies in particular that the wave functionals give approximately disjoint magnetic field distributions. Since different magnetic fields $B_i$ correspond to different fields $A^T_i$, we have that these wave functionals also give approximately disjoint distributions for the fields $A^T_i$.

This feature is important since it implies that, in the context of a more complete theory which also includes a description of matter, like for example quantum electrodynamics, measurement results can get recorded and displayed in the field beable $A^T_i$. This observation was used in \cite{struyve06,struyve07c} to argue that it is actually sufficient to have only beables for the electromagnetic field in order to have a pilot-wave model that reproduces the predictions of, say, quantum electrodynamics. We discuss this approach further in section \ref{minimalist}.

In section \ref{someelementarystates}, we considered some particular wave functionals for the Schr\"odinger field and the corresponding field distributions. The results of that discussion also apply to the case of the electromagnetic field. Wave functionals corresponding to a single photon will in general have significant overlap. On the other hand, coherent states that are ``sufficiently distinct'' are non-overlapping. With ``sufficiently distinct'' we mean for example that the states correspond to a different average photon number or that they correspond to photons with different momenta. For wave functionals corresponding to a definite but high number of photons it is unclear whether they can be non-overlapping. Note that a macroscopic state for the electromagnetic field is not a state with a definite number of particles (whereas we consider macroscopic states for ordinary matter to be states with an approximately definite number of particles), since for these states the expectation values of electric and magnetic field operators are zero. Instead a macroscopic state contains, just as coherent states (apart from the ground state), an indefinite number of particles.

\subsection{Valentini's approach: equivalence class of potentials as beable}\label{valentinisapproach}
Valentini has a different approach to gauge theories \cite{valentini92,valentini96,valentini09}. But while the ontology is different than that in Bohm's theory, the theories appear to be empirically equivalent, even in non-equilibrium.

Valentini's pilot-wave model is best understood in the context of the quantization procedure in which constraints are imposed as conditions on states. Therefore we start with recalling this quantization procedure for the electromagnetic field.

\subsubsection{Quantization by imposing the constraints as conditions on states}
Instead of quantizing only gauge independent degrees of freedom, the electromagnetic field can also be quantized as follows (cf.\ section \ref{canonicalquantization}). All canonical variables are quantized as if there were no constraints, i.e.\ 
\begin{equation}
[{\widehat A}_{0}({\bf x}),{\widehat \Pi}_{A_{0}}({\bf y})] = \ii \delta({\bf x} - {\bf y})\,, \quad {[{\widehat A}_{i}({\bf x}),{\widehat \Pi}_{A_{j}}({\bf y})]} = \ii \delta_{ij}\delta({\bf x} - {\bf y})\,,
\label{e.23}
\end{equation}
where the other fundamental commutation relations are zero, and the constraints are imposed as conditions on states, i.e.\
\begin{equation}
{\widehat \chi}_1 |\Psi \rangle = {\widehat \Pi}_{A_{0}}|\Psi \rangle = 0, \quad {\widehat \chi}_2 |\Psi \rangle =\partial_i {\widehat \Pi}_{A_{i}}|\Psi \rangle = 0\,.  
\label{e.24}
\end{equation}
Because the operators satisfy the standard commutation relations, the functional Schr{\"o}\-ding\-er representation can be used:
\begin{align}
{\widehat A}_{0}({\bf x}) &\to A_{0} ({\bf x}),\quad {\widehat \Pi}_{A_{0} }({\bf x}) \to -\ii\frac{\delta}{\delta A_{0} ({\bf x})}\,, \nonumber\\
{\widehat A}_{i}({\bf x}) &\to A_{i} ({\bf x}),\quad {\widehat \Pi}_{A_{i} }({\bf x}) \to -\ii\frac{\delta}{\delta A_{i} ({\bf x})}\,.
\label{e.25}
\end{align}
The corresponding functional Schr\"o\-ding\-er equation for the wave\-func\-tion\-al $\Psi(A_{0},A_i,t)=\la A_{0},A_i|\Psi(t)\ra$ reads 
\begin{equation}
\ii\frac{\partial \Psi}{\partial t}= \int d^3 x \left( -\frac{1}{2}\frac{\delta^2}{\delta  A_{i}\delta  A_{i}} + \frac{1}{4}F_{ij}F_{ij}  \right)\Psi 
\label{e.26}
\end{equation}
and physical states further have to satisfy the constraints ({\ref{e.24}}), i.e.\ 
\begin{align}
\frac{\delta \Psi}{\delta  A_{0}}&=0\,, \label{e.27}\\
\partial_i \frac{\delta \Psi}{\delta  A_{i}}&=0\,.
\label{e.28}
\end{align}

The quantum field theory so obtained is equivalent to the quantum field theory that was obtained by quantizing only the gauge independent degrees of freedom. This can be seen as follows. The first constraint ({\ref{e.27}}) simply states that $\Psi$ does not depend on $A_0$. The second constraint ({\ref{e.28}}) further implies that 
\begin{equation}
\frac{\delta \Psi}{\delta A^L_i} =  \frac{\partial_i \partial_j }{\nabla^2} \frac{\delta \Psi }{\delta A_j} = 0\,,
\label{e.29}
\end{equation}
i.e.\ physical states should also not depend on the longitudinal part of the vector potential. Hence physical states are of the form $\Psi( A^T_i,t)$. For these states the functional Schr{\"o}\-ding\-er equation ({\ref{e.26}}) reduces to 
\begin{equation}
\ii\frac{\partial \Psi}{\partial t} = \frac{1}{2}  \int d^3 x \left( -\frac{\delta^2}{ \delta A^T_i A^T_i} - A^T_i \nabla^2 A^T_i  \right)\Psi \,,
\label{e.30}
\end{equation}
which is the Schr{\"o}\-ding\-er equation ({\ref{e.20}}) that was obtained by quantizing only the gauge independent degrees of freedom. 

In order to obtain full equivalence of the quantum field theories, one also needs to show the equivalence of the inner product. As mentioned in section \ref{canonicalquantization}, the definition of the inner product requires some care in the case constraints are imposed as conditions on states. In the Hilbert space of unconstrained wave functionals $\Psi(A_0,A_i)$, one can define the inner product as
\begin{equation}
\la \Psi_1 | \Psi_2 \ra = \int  {\mathcal D} A_0 \left(\prod_i {\mathcal D} A_i\right) \Psi^*_1 \Psi_2\,. 
\label{e.31}
\end{equation}
However, physical states do not depend on $A_0$ and $A^L_i$ and therefore for physical states the expression $\la \Psi_1 | \Psi_2 \ra$ becomes proportional to the infinite factor $\int {\mathcal D} A_0  {\mathcal D} A^L$, the so-called ``gauge volume'', which makes the wave functionals non-normalizable. The solution is to factor out this infinite gauge volume from the expression $\la \Psi_1 | \Psi_2 \ra$ and to define the finite remnant as the inner product for physical states, i.e.\
\begin{equation}
\la \Psi_1 | \Psi_2 \ra_{ph} = \int {\mathcal D} A^T \Psi^*_1 \Psi_2\,. 
\label{e.32}
\end{equation}
This inner product is the same as the one used in the case where only the gauge independent degrees of freedom of the electromagnetic field are quantized.

In general the gauge volume can be factored out by using the Faddeev-Popov procedure. This procedure basically consists in inserting the quantity $\delta(f(A_0,A_i))\Delta_{FP,f}(A_0,A_i)$ in the integral in \eqref{e.31}, where the delta function imposes a gauge fixing (i.e.\ the condition $f(A_0,A_i)=0$ selects a unique representant from each set of gauge equivalent fields) and where $\Delta_{FP,f}$ is the corresponding Faddeev-Popov determinant which ensures that this procedure is independent of the choice of gauge. In the case of the electromagnetic field there are many suitable gauge fixings. For example the form \eqref{e.32} for the inner product is readily obtained by choosing the gauge $\partial_i A_i = 0$, $A_0=0$. In the case of non-Abelian Yang-Mills theories it is more difficult to apply the Faddeev-Popov procedure, because a large class of possible gauge fixings is ruled out \cite{gribov78,singer78,chodos80}. Therefore, for non-Abelian Yang-Mills theories, the Faddeev-Popov procedure is often applied only locally, which is fine within a perturbative treatment.

\subsubsection{Pilot-wave model}
Valentini did not really follow the above quantization scheme to arrive at his pilot-wave model. Instead he started by arguing that the degree of freedom $A_0$ in the theory of the classical electromagnetic field is a mathematical artifact, arising from the insistence on Lorentz invariance. Valentini therefore dropped the degree of freedom $A_0$ from the outset. In the resulting quantum theory, the quantum state $\Psi$ is a wave functional defined on the space of vector fields $A_i$ which satisfies the functional Schr{\"o}\-ding\-er equation
\begin{equation}
\ii \frac{\partial \Psi}{\partial t}=  \int d^3 x \left( -\frac{1}{2}\frac{\delta^2}{\delta  A_{i}\delta  A_{i}} + \frac{1}{4}F_{ij}F_{ij}  \right)\Psi \,,
\label{64.003}
\end{equation}
together with the constraint
\begin{equation}
\partial_i \frac{\delta \Psi}{\delta  A_{i}} =0\,.
\label{64.004}
\end{equation}
Valentini obtained the constraint by requiring that the wave functional be invariant under infinitesimal time-independent gauge transformations of the form
\begin{equation}
A_i \to A_i + \partial_i \theta\,.
\label{64.005}
\end{equation}

This quantum field theory can also be obtained by following the scheme presented above. The theory can be seen to follow from imposing the constraints as conditions on quantum states, by solving the constraint ({\ref{e.27}}) but keeping the constraint ({\ref{e.28}}). Alternatively it can be found by imposing the temporal gauge $A_0=0$, see e.g.\ \cite[p.\ 398]{jackiw95}.

In his pilot-wave approach Valentini introduced a field $A_i$ for which the dynamics is determined by the guidance equation
\begin{equation}
{\dot A}_{i}  = \frac{\delta S}{\delta A_i } \,.
\label{74}
\end{equation}
The velocity field is the one that enters in the continuity equation for $|\Psi|^2$:
\begin{equation}
\frac{\partial |\Psi|^2}{\partial t}  +   \int d^3x  \frac{\delta }{\delta  A_{i} }  \left(|\Psi|^2 \frac{\delta S}{\delta A_i } \right) =0\,. 
\label{71}
\end{equation}
However, instead of regarding the field $A_i$ as the beable, Valentini suggests to regard what he calls a ``Faraday geometry'' as the actual beable, with a Faraday geometry being an element of what he calls ``electrodynamic superspace'' \cite[p.\ 73]{valentini92}. A ``Faraday geometry'' is hereby understood as an equivalence class of fields connected by the gauge transformations \eqref{64.005} and the ``electrodynamic superspace'' is the space of such equivalence classes. This is analogous to Wheeler's notion of ``superspace'' in geometrodynamics \cite{wheeler64,wheeler68}, which is the space of 3-geometries, a 3-geometry being an equivalence class of 3-metrics connected by spatial diffeomorphisms. 

Let us consider the ontology in more detail. First note that that the constraint ({\ref{64.004}}) is solved by wave functionals that only depend on the transverse part of the vector potential and that such wave functionals obey the functional Schr{\"o}\-ding\-er equation ({\ref{e.20}}). As such, Valentini's guidance equation \eqref{74} reduces to 
\begin{equation}
{\dot A}^T_{i}  = \frac{\delta S}{\delta A^T_i } \,, \qquad {\dot A}^L_{i}  = \frac{\delta S}{\delta A^L_i } = 0 \,. 
\label{74.002} 
\end{equation}
Because the wave functional $\Psi( A^T_i,t)$ satisfies the same functional Schr{\"o}\-ding\-er equation as in Bohm's pilot-wave model, the dynamics of the transverse field is the same as in Bohm's approach. The longitudinal field is static and has no effect on the dynamics of the transverse field. Hence, the longitudinal field appears to be an unphysical degree of freedom. Valentini's suggestion to take the equivalence class of fields $A_i$ as the actual beable in effect eliminates this unphysical degree of freedom by treating all its possible values as physically equivalent.{\footnote{Note that the dynamics of the field $A_i$ (including the field $A^L_i$) is uniquely determined by the equations of motion and the initial conditions. Therefore, strictly speaking, this pilot-wave theory does not constitute a gauge theory in the sense described in section \ref{canonicalquantization}. Hence, although the transformation $A_i \to A_i + \partial_i \theta$ is a symmetry of the theory, it is is not a gauge symmetry in this sense. Nevertheless, as discussed above $A^L_i$ is clearly an unphysical degree of freedom and can be dismissed on this basis.}} As such, it is clear that, while the ontology in Valentini's approach is strictly speaking different from that in Bohm's approach, they are empirically equivalent, even in quantum non-equilibrium. 

Of course there exist other ways to eliminate the gauge symmetry \eqref{64.005}. For example, one could choose a unique representant from each equivalence class, or one could choose an invariant function of the field $A_i$, such as the field $B_i = \epsilon_{ijk} \partial_j A_k$. While all these choices correspond to different ontologies, they would lead to empirically equivalent theories, even in quantum non-equilibrium. Such a type of ambiguity does not only appear for pilot-wave theories, but is common for gauge theories. For example, in classical electromagnetism different possible ontologies are possible that are empirically equivalent, such as for example an ontology given by the electric and magnetic field or an ontology given by gauge-fixed potential (see \cite{healey07} for an extensive discussion). 

We finish this section with some notes on the quantum equilibrium distribution in Valentini's approach. From ({\ref{74.002}}), it is clear that an equivalence class will evolve to an equivalence class. A measure on the set of transverse fields naturally determines a measure on the set of equivalence classes. Hence the quantum equilibrium measure $|\Psi(A^T_i,t)|^2 {\mathcal D} A^T$ in Bohm's pilot-wave approach naturally defines a quantum equilibrium measure on the set of equivalence classes. More generally, one could define the equilibrium measure using the Faddeev-Popov procedure. Given a suitable gauge fixing $f(A_i)=0$, one has the measure $|\Psi(A_i,t)|^2 \delta (f(A_i)) \Delta_{FP,f}(A_i) \prod_j{\mathcal D} A_j$ on the space of fields $A_i$, which defines a natural measure on the space of equivalence classes. The resulting measure is independent of the choice of gauge fixing, so that, in particular, it agrees with the one defined in terms of transverse fields. 

In the following sections we will discuss scalar quantum electrodynamics and the Abelian Higgs model. We will present a pilot-wave model for these field theories by introducing beables only for gauge independent degrees of freedom, thereby avoiding the surplus structure in Valentini's approach. For Valentini's approach to scalar quantum electrodynamics, see \cite{valentini92}. For his approach to the Abelian Higgs model, as well as Yang-Mills theories, see \cite{valentini09}. It is expected that the equivalence between Valentini's approach and an approach in terms of gauge independent degrees of freedom, which was established here in the case of the free electromagnetic field, can be established much more generally.


\section{Scalar quantum electrodynamics}\label{scalarqed}
In this section we present a pilot-wave model for scalar quantum electrodynamics (a quantized scalar field interacting with a quantized electromagnetic field). The model is obtained by introducing beables for gauge independent degrees of freedom, just as in Bohm's pilot-wave model for the free electromagnetic field (for Valentini's pilot-wave approach to scalar quantum electrodynamics, see \cite{valentini92}).

The Lagrangian and the corresponding equations of motion for classical scalar electrodynamics are given by
\begin{equation}
L =  \int d^3 x \left((D_{\mu} \phi)^* D^{\mu} \phi - m^2\phi^*\phi - V(\phi^*\phi) - \frac{1}{4} F^{\mu \nu}F_{\mu \nu} \right) \,,
\label{sq.1}
\end{equation}
\begin{equation}
D_{\mu}D^{\mu}\phi + m^2 \phi + \frac{\delta V}{\delta\phi^* }= 0\,,\quad D^*_{\mu}D^{\mu*} \phi^* + m^2 \phi^* + \frac{\delta V}{\delta\phi } = 0\,,
\label{sq.2}
\end{equation}
\begin{equation}
\partial_{\mu} F^{\mu \nu}= s^{\nu} = \ii e\left(\phi^*D^{\nu}\phi - \phi D^{\nu*} \phi^* \right)\,,
\label{sq.3}
\end{equation}
with $D_{\mu}= \partial_{\mu} + \ii e A_{\mu}$ the covariant derivative and $s^{\mu}$ the charge current. The theory has a local $U(1)$ symmetry
\begin{equation}
\phi \to e^{\ii e \theta} \phi \,, \quad \phi^* \to e^{-\ii e \theta} \phi^*\,,\quad  A^{\mu} \to   A^\mu -  \partial^\mu  \theta\,, 
\label{sq.4}
\end{equation}
which signals the presence of gauge freedom.{\footnote{Unlike the case of free electromagnetism, not all of the transformations \eqref{sq.4} form gauge transformations in the sense defined in section \ref{canonicalquantization}. Writing $g=e^{\ii e \theta}$ the local $U(1)$ transformation \eqref{sq.4} becomes $\phi \to g \phi$, $\phi^* \to g^{-1} \phi^*$, $A^{\mu} \to   A^\mu + i g^{-1}  \partial^\mu g/e$. Assuming as boundary conditions that the fields $A_\mu$ vanish sufficiently fast at spatial infinity, the only functions $g$ that preserve these boundary conditions are those that have a limit in $U(1)$ at spatial infinity and reach that limit sufficiently fast. One can easily show that, of those transformations, only those that go to one at spatial infinity are gauge transformations \cite{struyve09c}. Modding out over the gauge transformations yields a residual global $U(1)$ symmetry. This residual symmetry is then present when the theory is expressed in terms of gauge independent variables.\label{sqedgauge}}} Hence, just as in the case of the free electromagnetic field there will be first class constraints in the Hamiltonian picture.

Indeed, in the Hamiltonian picture we have \cite[pp.\ 113-127]{gitman90}:
\begin{equation}
\Pi_{\phi} = \frac{\delta L }{\delta {\dot \phi}  } = D^*_0 \phi^{*} , \quad \Pi_{\phi^*} = \frac{\delta L }{\delta {\dot \phi^*}  } = D_0 \phi \,, 
\label{sq.5}
\end{equation}
\begin{equation}
\Pi_{A_0} = \frac{\delta L }{\delta {\dot A_0} } = 0, \quad \Pi_{A_i} = \frac{\delta L }{\delta {\dot A_i} } = ({\dot A}_i +\partial_i A_0 )\,,
\label{sq.6}
\end{equation}
\begin{eqnarray}
H &=&   \int d^3 x \bigg(\Pi_{\phi^*} \Pi_{\phi} + \left(D_{i} \phi \right)^{*}  D_{i} \phi + m^2 \phi^*\phi + V(\phi^*\phi)\nonumber\\
&&  \mbox{} +  \frac{1}{2} \Pi_{A_i}\Pi_{A_i} + \frac{1}{4} F_{ij}F_{ij} + A_0\left( \partial_i \Pi_{A_i} + s_0 \right) \bigg)\,,
\label{sq.7}
\end{eqnarray}
\begin{equation}
\chi_{1} = \Pi_{A_0}\,, \quad \chi_{2} = \partial_i \Pi_{A_i} + s_0 =  \partial_i \Pi_{A_i} + \ii e\left(\phi^*  \Pi_{\phi^*} - \phi \Pi_{\phi} \right) \,,
\label{sq.8}
\end{equation}
where the constraints $\chi_1$ and $\chi_2$ are first class.

Since $\Pi_{A_0}$ is a constraint, the variable $A_0$ can, just as in the case of the free electromagnetic field, immediately be identified as a gauge variable. In order to identify the remaining gauge variables we first perform the following canonical transformation 
\begin{align}
\phi &= e^{\ii e \frac{\partial_i}{\nabla^2} {\bar A}_i}  {\bar \phi}\,, & \Pi_{\phi} &= e^{-\ii e \frac{\partial_i}{\nabla^2} {\bar A}_i} \Pi_{{\bar \phi}}\,, \nonumber\\
\phi^* &= e^{-\ii e \frac{\partial_i}{\nabla^2} {\bar A}_i}  {\bar \phi}^*\,, & \Pi_{\phi^*} &= e^{\ii e \frac{\partial_i}{\nabla^2} {\bar A}_i} \Pi_{{\bar \phi}^*}\,, \nonumber\\
A_i &= {\bar A}_i\,, & \Pi_{A_i} &= \Pi_{{\bar A}_i} - \frac{\partial_i}{\nabla^2} {\bar s}_0\,,
\label{sq.9}
\end{align}
where ${\bar s}_0 = \ii e\left({\bar \phi}^*   \Pi_{{\bar \phi}^*} - {\bar \phi}  \Pi_{{\bar \phi} } \right)$ is the charge density expressed in terms of the new variables. The Hamiltonian for the new variables is given by{\footnote{Note that, as alluded before in footnote \ref{sqedgauge}, there is a residual global $U(1)$ symmetry, namely ${\bar \phi} \to g{\bar \phi}$, ${\bar \phi}^* \to g^{-1}{\bar \phi}^*$, $\Pi_{{\bar \phi}} \to g^{-1}\Pi_{{\bar \phi}}$, $\Pi_{{\bar \phi}^*} \to g\Pi_{{\bar \phi}^*}$, with $g$ a constant phase factor.}}
\begin{multline}
H = \int d^3 x \bigg(\Pi_{{\bar \phi}^*} \Pi_{{\bar \phi}} + \left(D^T_{i} {\bar \phi}  \right)^{*}  D^T_{i} {\bar \phi}  + m^2 {\bar \phi}^* {\bar \phi} +  V({\bar \phi}^*{\bar \phi}) -\frac{1}{2} {\bar s}_0  \frac{1}{\nabla^2} {\bar s}_0   \\
 +  \frac{1}{2} \Pi_{{\bar A}_i}\Pi_{{\bar A}_i} +  \frac{1}{4} {\bar F}_{ij}{\bar F}_{ij}  + A_0  \partial_i \Pi_{{\bar A}_i}\bigg)\,, \label{108}
\end{multline}
where 
\begin{equation}
D^T_{i} = \partial_{i} - \ii e\left(\delta_{ij} - \frac{\partial_i \partial_j}{\nabla^2} \right) {\bar A}_j =  \partial_{i} - \ii e{\bar A}^T_i
\label{sq.10}
\end{equation}
only depends on the transverse degrees of freedom of ${\bar A}_i$ and ${\bar F}_{ij} = \partial_i {\bar A}_j - \partial_j {\bar A}_i$. The term containing ${\bar s}_0$ is the Coulomb energy.

In terms of the new canonical variables the remaining constraint $\chi_{2} = \partial_i \Pi_{A_i} + s_0 =0$ reads $\partial_i \Pi_{{\bar A}_i}=0$. So the remaining constraint now has the same form as in the case of the free electromagnetic field. By performing a Fourier transformation on ${\bar A}_i$ and $\Pi_{{\bar A}_i}$, similarly as in the case of the free electromagnetic field, cf.\ (\ref{e.9}), the remaining gauge degree of freedom can be separated from the gauge independent degrees of freedom. The quantization then proceeds in the same way as in the case of the free electromagnetic field. 

The fact that the fields ${\bar A}^T_i,{\bar \phi},{\bar \phi}^*$ and the corresponding conjugate fields are gauge independent variables is well known, see e.g.\ \cite[pp.\ 302-306]{dirac67}, and motivated the canonical transformation (\ref{sq.9}). The resulting theory can also be obtained by imposing the Coulomb gauge.

After quantization and by using the functional Schr\"odinger picture, we obtain the functional Schr\"odinger equation 
\begin{eqnarray}
\ii\frac{\partial \Psi}{\partial t} &=&   \int d^3 x \bigg(-\frac{ \delta^2  }{ \delta {\bar \phi}^* \delta {\bar \phi}} + \left(D^T_{i} {\bar \phi}  \right)^{*}  D^T_{i} {\bar \phi}  + m^2 {\bar \phi}^* {\bar \phi} + V({\bar \phi}^*{\bar \phi})   \nonumber\\
&& \mbox{}- \frac{e^2}{2} \left({\bar \phi}^* \frac{ \delta  }{ \delta {\bar \phi}^*} -  {\bar \phi} \frac{ \delta }{ \delta {\bar \phi}} \right) \frac{1}{\nabla^2} \left({\bar \phi}^* \frac{ \delta  }{ \delta {\bar \phi}^*} -  {\bar \phi} \frac{ \delta }{ \delta {\bar \phi}} \right) \nonumber\\
&& \mbox{}- \frac{1}{2} \frac{\delta^2}{ \delta {\bar A}^T_i {\bar A}^T_i} -  \frac{1}{2} {\bar A}^T_i \nabla^2 {\bar A}^T_i \bigg)\Psi \,,
\label{sq.11}
\end{eqnarray}
for the wave functional $\Psi({\bar \phi},{\bar \phi}^*,{\bar A}^T_i,t)$. Beables ${\bar \phi},{\bar \phi}^*,{\bar A}^T_i $ can be introduced which evolve according to the guidance equations
\begin{eqnarray}
{\dot{\bar \phi} }  &=&  \frac{\delta S}{\delta {\bar \phi}^* }  + e^2  {\bar \phi} \frac{1}{\nabla^2} \left(  {\bar \phi}\frac{\delta S}{\delta {\bar \phi} } - {\bar \phi}^*\frac{\delta S}{\delta {\bar \phi}^* } \right) \,,  \nonumber\\ 
{\dot{\bar \phi} }^*  &=&  \frac{\delta S}{\delta {\bar \phi} }  + e^2  {\bar \phi}^* \frac{1}{\nabla^2} \left( {\bar \phi}^*\frac{\delta S}{\delta {\bar \phi}^* }  - {\bar \phi}\frac{\delta S}{\delta {\bar \phi} } \right) \,,  \nonumber\\ 
{\dot{\bar A}^T_i } &=&  \frac{\delta S}{\delta {\bar A}^T_i  }
\label{sq.12}
\end{eqnarray}
and for which the quantum equilibrium density is given by $|\Psi({\bar \phi},{\bar \phi}^*,{\bar A}^T_i,t)|^2$.


\section{Massive spin-1 field}
In this section we present a pilot-wave model for a massive spin-1 field. The reason to do this, is that we will encounter a massive spin-1 field when we discuss the Abelian Higgs model in the next section. The pilot-wave model was presented before by Takabayasi \cite{takabayasi52}.

The Lagrangian for a classical spin-1 field with mass $m$ is obtained by adding a mass term to the Lagrangian for a classical massless spin-1 field (which is just the Lagrangian for the electromagnetic field)
\begin{equation}
L = \int d^3 x \left( -\frac{1}{4} F^{\mu \nu}F_{\mu \nu} + \frac{m^2}{2}  A^\mu A_\mu  \right) \,. 
\label{ms.1}
\end{equation}
The corresponding equation of motion is 
\begin{equation}
\partial_\mu F^{\mu \nu} = - m^2 A^\nu \,. 
\label{ms.2}
\end{equation}

The Hamiltonian formulation is determined by \cite[p.\ 35]{gitman90}
\begin{equation}
\Pi_{A_0} = \frac{\delta L }{\delta {\dot A_0} } = 0\,, \quad \Pi_{A_i} = \frac{\delta L }{\delta {\dot A_i} } = ({\dot A}_i +\partial_i A_0 )\,,
\label{ms.3}
\end{equation}
\begin{equation}
H = \int d^3 x \left( \frac{1}{2}\Pi_{A_i}\Pi_{A_i} + \frac{1}{4} F_{ij}F_{ij} + \frac{m^2}{2} A_iA_i - \frac{m^2}{2} A^2_0+A_0\partial_i \Pi_{A_i} \right) \,,
\label{ms.4}
\end{equation}
\begin{equation}
\chi_{1} = \Pi_{A_0} \,, \quad \chi_{2} = \partial_i \Pi_{A_i} - m^2A_0\,,
\label{ms.5}
\end{equation}
where $\chi_{1}$ and $\chi_{2}$ are second class constraints. By performing the canonical transformation
\begin{eqnarray}
A_0 &=& {\bar A}_0 +  \frac{1}{m^2} \pa_i \Pi_{{\bar A}_i} \,, \quad \Pi_{A_0} = \Pi_{{\bar A}_0} \,, \nonumber\\
A_i &=& {\bar A}_i +  \frac{1}{m^2} \pa_i \Pi_{{\bar A}_0}  \,, \quad \Pi_{A_i} = \Pi_{{\bar A}_i} \,,
\label{ms.6}
\end{eqnarray}
we find that the constraints reduce to ${\bar A}_0=\Pi_{{\bar A}_0}=0$. The unconstrained variables are ${\bar A}_i$ and $\Pi_{{\bar A}_i}$. The Hamiltonian for these variables reads
\begin{equation}
H = \int d^3 x \left( \frac{1}{2}\Pi_{{\bar A}_i}\Pi_{{\bar A}_i} + \frac{1}{2m^2}\left( \partial_i \Pi_{{\bar A}_i}  \right)^2  + \frac{1}{4} {\bar F}_{ij}{\bar F}_{ij}  +  \frac{m^2}{2}  {\bar A}_i {\bar A}_i \right)\,.
\label{ms.7}
\end{equation}

After quantization and by using the functional Schr{\"o}\-ding\-er picture, we obtain the functional Schr{\"o}\-ding\-er equation
\begin{equation}
\ii\frac{\partial \Psi}{\partial t}= \int d^3 x \left( -\frac{1}{2}\frac{\delta^2}{\delta  {\bar A}_{i}\delta  {\bar A}_{i}}-  \frac{1}{2m^2}\left( \partial_i  \frac{\delta}{\delta  {\bar A}_{i} } \right)^2  + \frac{1}{4}{\bar F}_{ij}{\bar F}_{ij} +  \frac{m^2}{2}  {\bar A}_i {\bar A}_i  \right)\Psi\,.
\label{115.004}
\end{equation}
The guidance equation for the field beable ${\bar A}_{i}$ reads
\begin{equation}
{\dot {\bar A}}_{i}  =  \frac{\delta S}{\delta {\bar A}_i } - \frac{1}{m^2} \partial_i\partial_j \frac{\delta S}{\delta {\bar A}_j }\,.
\label{115.006}
\end{equation}

\section{Abelian Higgs model}
The electromagnetic interaction is unified with the weak interaction in the electroweak theory of Glashow, Salam and Weinberg. Instead of introducing beables for the electromagnetic field, it might therefore be more fundamental to introduce beables for the electroweak field. However, the electroweak theory is a non-Abelian gauge theory, based on the symmetry group $SU(2) \times U(1)_Y$. Due to the technical difficulties that come with the construction of a pilot-wave model for non-Abelian gauge theories (cf.\ section \ref{nonabeliangaugetheories}), we do not make an attempt to construct a pilot-wave model for the electroweak theory.

However, there is one important aspect of the electroweak theory that we want to discuss in the context of pilot-wave theory, namely spontaneous symmetry breaking and the associated Higgs mechanism. In the electroweak theory, the $SU(2) \times U(1)_Y$ symmetry gets broken to the $U(1)_{em}$ symmetry of electromagnetism. In this process the weak interaction bosons $W^+,W^-,Z^0$ acquire mass through the Higgs mechanism. That the pilot-wave theory has no problem in dealing with spontaneous symmetry breaking and the Higgs mechanism can be illustrated in the simple case of the Higgs model (which has a $U(1)$ symmetry) (for Valentini's pilot-wave approach see \cite{valentini09}).

Before turning to quantum field theory, let us first consider the Abelian Higgs model on the level of classical field theory. The Abelian Higgs model (see e.g.\ \cite{abers73}) describes a scalar field coupled to an electromagnetic field, for which the classical equations of motion can be derived from the Lagrangian
\begin{equation}
L =  \int d^3 x \left((D_{\mu} \phi)^* D^{\mu} \phi   + \mu^2\phi^*\phi - \lambda(\phi^*\phi)^2 - \frac{1}{4} F^{\mu \nu}F_{\mu \nu} \right) \,.
\label{h.1}
\end{equation}
The constant $\mu$ is real, so that the mass of the scalar field is imaginary, making the field tachyonic. The constant $\lambda$ is real and positive. This Lagrangian is the one of scalar electrodynamics, cf.\ section \ref{scalarqed}, with potential $V(\phi^*\phi)=\lambda(\phi^*\phi)^2$. The theory exhibits a local $U(1)$ symmetry, given by the transformations in (\ref{sq.4}).

Broadly speaking, one has spontaneous symmetry breaking if the ground state (the state of lowest energy) does not share the symmetry of theory, but is degenerate in such a way that the different ground states are mapped to each other by the symmetry. To each ground state there corresponds a set of solutions to the equations of motion that arise as perturbations around that ground state. The effective equations of motion for each of those sets generically do not have the symmetry of the original equations of motion.

With the spontaneous symmetry breaking of a gauge symmetry new features arise, such as the generation of mass through the Higgs mechanism. However, since gauge symmetries are unphysical, some care is required with the meaning of symmetry breaking in this case. This will be considered in detail in \cite{struyve09c}. Strictly speaking, the symmetry that is broken in the case of the Abelian Higgs model is not the gauge symmetry but the residual global $U(1)$ symmetry that arises after modding out the local $U(1)$ symmetries by the actual gauge symmetries (see footnote \ref{sqedgauge}). 

One way to see this is by considering the formulation of the theory in terms of gauge invariant variables for which the Hamiltonian is given by{\footnote{Usually the analysis of symmetry breaking in the Abelian Higgs mechanism is carried out by employing the unitary gauge.}}
\begin{multline}
H = \int d^3 x \bigg(\Pi_{{\phi}^*} \Pi_{ \phi} + \left(D^T_{i}  \phi  \right)^{*}  D^T_{i}  \phi -  \mu^2\phi^*\phi + \lambda(\phi^*\phi)^2   -\frac{1}{2}  s_0  \frac{1}{\nabla^2} s_0   \\
 +  \frac{1}{2} \Pi_{A^T_i}\Pi_{A^T_i} - \frac{1}{2} A^T_i \nabla^2 A^T_i \bigg)\,, 
\label{h.2}
\end{multline}
where we dropped the bars on top of the fields for notational simplicity. 

In this theory the gauge symmetry has been eliminated, but, as noted before, it possesses a residual global $U(1)$ symmetry
\begin{equation}
 \phi \to e^{\ii e \theta}  \phi , \quad  \phi^* \to e^{-\ii e \theta} \phi^*,\quad \Pi_{ \phi} \to e^{-\ii e \theta}\Pi_{\phi} ,\quad \Pi_{ \phi^*} \to e^{\ii e \theta} \Pi_{ \phi^*}. 
\label{h.3}
\end{equation}
The minima of the energy, i.e.\ the Hamiltonian, are given by $\phi = \sqrt{\mu^2/2\lambda} e^{\ii \alpha}$, $\phi^* = \sqrt{\mu^2/2\lambda} e^{-\ii \alpha}$  (where $\alpha$ is constant) and $\Pi_{ \phi}=\Pi_{ \phi^*}=A^T_i=\Pi_{A^T_i}=0$. Let us now choose one of those ground states, say $ \phi = \sqrt{\mu^2/2\lambda}=v/\sqrt{2}$, and make the following change of variables (which forms a canonical transformation)
\begin{align}
 \phi &= \frac{1}{\sqrt{2}} (v + \eta + \ii \xi) , & \Pi_{ \phi} &= \frac{1}{\sqrt{2}} (\Pi_{\eta}  - \ii \Pi_{\xi} ), \nonumber\\
 \phi^* &= \frac{1}{\sqrt{2}} (v + \eta - \ii \xi) , & \Pi_{ \phi^*} &= \frac{1}{\sqrt{2}} (\Pi_{\eta}  + \ii \Pi_{\xi} ),
\label{h.4}
\end{align}
where $\eta,\Pi_{\eta},\xi,\Pi_{\xi}$ are real variables. Let us also introduce
\begin{equation}
A^L_i = - \frac{1}{ev}\partial_i \xi , \quad \Pi_{A^L_i} = ev \frac{\partial_i}{\nabla^2}\Pi_{\xi}.
\label{h.5}
\end{equation}
This transformation is not canonical; the Poisson bracket of $A^L_i$ and $\Pi_{A^L_i}$ is given by the longitudinal delta function. The latter variables allow us to introduce the variables 
\begin{equation}
A_i =A^T_i +A^L_i , \quad \Pi_{A_i} = \Pi_{A^T_i} + \Pi_{A^L_i}.
\label{h.6}
\end{equation}
The variables $A_i, \Pi_{A_i}$ form canonical pairs again.

Expressing the Hamiltonian in terms of the variables $\eta,\Pi_{\eta},\xi,\Pi_{\xi},A_i, \Pi_{A_i}$, only up to quadratic terms and up to a (infinite) constant, we obtain
\begin{multline}
H = \int d^3 x \bigg(\frac{1}{2} \Pi^2_{\eta}  + \frac{1}{2} \partial_i \eta \partial_i \eta + \mu^2 \eta^2 \\
+ \frac{1}{2} \Pi_{A_i}\Pi_{A_i} + \frac{1}{2e^2v^2} \left(\partial_i \Pi_{A_i}\right)^2 + \frac{e^2v^2}{2} A_iA_i + \frac{1}{4}F_{ij}F_{ij}  \bigg). 
\label{h.7}
\end{multline}
The higher order terms contain the coupling between the fields. The first line is recognized as the Hamiltonian of a scalar field with mass $\sqrt{2\mu^2}$ and the second line as that of a spin-1 field with mass $\sqrt{e^2v^2}$. What has happened is that the Goldstone boson $\xi$ and the transverse field $A^T_i$ together form a massive spin-1 field. This is the Higgs mechanism. 

So far the discussion of the symmetry breaking and the Higgs mechanism was done on the classical level. A similar picture arises in the pilot-wave approach to the corresponding quantum field theory. Since the Lagrangian of the Abelian Higgs model corresponds to the one of scalar quantum electrodynamics, the quantized theory and the corresponding pilot-wave model are those presented in section \ref{scalarqed}. The functional Schr\"o\-ding\-er equation reads
\begin{multline}
\ii\frac{\partial \Psi}{\partial t} =   \int d^3 x \bigg(-\frac{ \delta^2  }{ \delta { \phi}^* \delta { \phi}} + \left(D^T_{i} { \phi}  \right)^{*}  D^T_{i} { \phi}  - \mu^2 { \phi}^* { \phi} +  \lambda(\phi^*\phi)^2  \\
- \frac{e^2}{2} \left({ \phi}^* \frac{ \delta  }{ \delta { \phi}^*} -  { \phi} \frac{ \delta }{ \delta { \phi}} \right) \frac{1}{\nabla^2} \left({ \phi}^* \frac{ \delta  }{ \delta { \phi}^*} -  { \phi} \frac{ \delta }{ \delta { \phi}} \right) - \frac{1}{2} \frac{\delta^2}{ \delta { A}^T_i { A}^T_i} -  \frac{1}{2} { A}^T_i \nabla^2 { A}^T_i \bigg)\Psi 
\label{h.8}
\end{multline}
and the guidance equations are 
\begin{eqnarray}
{\dot{ \phi} }  &=&  \frac{\delta S}{\delta { \phi}^* }  + e^2  { \phi} \frac{1}{\nabla^2} \left(  { \phi}\frac{\delta S}{\delta { \phi} } - { \phi}^*\frac{\delta S}{\delta { \phi}^* } \right) \,,  \nonumber\\ 
{\dot{ \phi} }^*  &=&  \frac{\delta S}{\delta { \phi} }  + e^2  { \phi}^* \frac{1}{\nabla^2} \left( { \phi}^*\frac{\delta S}{\delta { \phi}^* }- { \phi}\frac{\delta S}{\delta { \phi} }  \right) \,,  \nonumber\\ 
{\dot{ A}^T_i } &=&  \frac{\delta S}{\delta { A}^T_i  } \,.
\label{h.9}
\end{eqnarray}

Let us now use the expansion $\phi=( v + \eta + \ii \xi)/\sqrt{2}$ (just as in \eqref{h.4}) in the functional Schr{\"o}\-ding\-er equation and in the guidance equations. We will write the functional Schr{\"o}\-ding\-er equation only up to quadratic terms in the fields $\eta,\xi$ and ${ A}^T_i$, and up to a (infinite) constant. The guidance equations will only be written up to linear terms in the fields. The higher order terms represent the interactions between the fields. By introducing the field ${ A}_i={ A}^T_i +{ A}^L_i $, where we define ${ A}^L_i = -\partial_i \xi/ ev$, and by making use of the relations
\begin{equation}
\frac{\delta }{\delta { \phi} } = \frac{1}{\sqrt{2}}\left( \frac{\delta }{\delta { \eta} } -\ii \frac{\delta }{\delta { \xi} } \right) \,,  \quad \frac{\delta }{\delta { A_i} } = \frac{\delta }{\delta { A^T_i} } + \frac{\delta }{\delta { A^L_i} }\,, \quad   \frac{\delta }{\delta { A^L_i} } = ev \frac{\pa_i}{\nabla^2} \frac{\delta }{\delta { \xi} }  \,,  
\label{h.10}
\end{equation}
we find
\begin{multline}
\ii\frac{\partial \Psi}{\partial t} = \int d^3 x  \bigg( - \frac{1}{2} \frac{\delta^2 }{\delta \eta^2 } +  \frac{1}{2} \partial_i {\eta} \partial_i { \eta}  +  \mu^2{ \eta}^2   - \frac{1}{2} \frac{\delta^2}{\delta  A_{i}\delta  A_{i}}  \\
-\frac{1}{2e^2v^2} \left( \partial_i  \frac{\delta}{\delta  A_{i} } \right)^2  + \frac{1}{4}F_{ij}F_{ij} + \frac{1}{2}e^2v^2  A_i A_i  \bigg)\Psi\,.
\label{h.11}
\end{multline}
Hence we recognize a scalar field $\eta$ with mass ${\sqrt{2\mu^2}}$ and a spin one field $A_i$ with mass ${\sqrt{e^2v^2}}$. The guidance equations reduce to 
\begin{equation}
{\dot \eta} =   \frac{\delta S}{\delta \eta }   \,, \quad {\dot A}_{i}  = \frac{\delta S}{\delta A_i } - \frac{1}{e^2v^2} \partial_i\partial_j \frac{\delta S}{\delta A_j } \,,
\label{h.12}
\end{equation}
which are the familiar guidance equations for a real scalar field and a massive spin-1 field.


\section{A note on non-Abelian gauge theories}\label{nonabeliangaugetheories}
For the development of a pilot-wave model for a gauge theory, our approach is to introduce only beables for gauge independent degrees of freedom. As explained before, the way to do this is by starting from the formulation of the quantum field theory which is obtained by quantizing only gauge independent degrees of freedom. For the gauge theories we have considered so far, this approach worked very well. One reason for this is that it was easy to identify and isolate the gauge independent degrees of freedom as canonical pairs. For non-Abelian Yang-Mills theories this is much harder. The reason is that the constraints are non-linear in the fields. For example for the free Yang-Mills field the constraints read $\Pi_{A^a_0} =0$ and $D_i \Pi_{A^a_i}= \pa_i \Pi_{A^a_i} - f_{abc} A^b_i\Pi_{A^c_i}=0$. These constraints look similar to the constraints for the free electromagnetic field, with the difference that the second constraint contains a non-linear term. Some progress has been made on the identification on gauge independent degrees of freedom for non-Abelian gauge fields, see e.g.\ \cite{karabali96} and references therein. This work could potentially be used for the development of a pilot-wave model.

In Valentini's approach \cite{valentini09}, where the actual beable is an equivalence class of gauge equivalent fields, no difficulties appear on the level of the dynamics, but instead on the level of specifying a quantum equilibrium distribution. In specifying the equilibrium distribution distribution, one needs to avoid the difficulties associated with the infinite gauge volume. As in the case of the electromagnetic field (cf.\ section \ref{valentinisapproach}), this could be done by applying the Faddeev-Popov procedure. However, it was shown that a large class of gauge fixings are ruled out \cite{singer78,chodos80}).


\section{Fermionic fields}\label{fermionicfields}
In this section we consider two pilot-wave approaches for fermionic fields that introduce fields as beables. One approach is by Holland \cite[pp.\ 449-457]{holland93b} and \cite{holland881}, the other by Valentini \cite{valentini92,valentini96}. We point out some problems with both these approaches.

\subsection{Holland's approach: Euler angles as beables}
Before turning to Holland's pilot-wave model for fermionic fields, we first have a look at a particular pilot-wave model for non-relativistic spin-1/2 particles, which was also proposed by Holland, cf.\ \cite[pp.\ 424-448]{holland93b} and \cite{holland881,holland882}. It is instructive to first consider this model because it bears some resemblance to his model for fermionic fields. The formulation of both models involves angular variables, but these variables play a significantly different role in each of the models. 

\subsubsection{Non-relativistic spin-1/2 particles}
The standard wave equation for non-relativistic spin-1/2 particles is the Pauli equation. Holland's starting point is a different wave equation which would underly the Pauli equation (although there is no evidence for such an underlying theory for elementary particles). 

For simplicity, we restrict our attention to a single spin-1/2 particle. In this case, the quantum state is given by a wave function $\psi({\bf x},{\boldsymbol \alpha},t)$, defined on the configuration space ${\mathbb R}^3 \times SU(2)$. As before, the coordinates ${\bf x}$ represent the position degree of freedom. The coordinates ${\boldsymbol \alpha}=(\alpha,\beta,\gamma)$, with $0 \leqslant \alpha \leqslant \pi$, $0 \leqslant \beta < 2\pi$, $0 \leqslant \gamma < 4\pi$, are Euler angles that parametrize $SU(2)$ and represent the spin degree of freedom.    

The group $SU(2)$ is introduced as the double covering group of the Euclidean rotation group $SO(3)$. The angles ${\boldsymbol \alpha}$ stand in two-to-one relation to a set of Euler angles parametrizing $SO(3)$, denoted by ${\boldsymbol \alpha}'=(\alpha',\beta',\gamma')$, with $0 \leqslant \alpha' \leqslant \pi$, $0 \leqslant \beta' < 2\pi$, $0 \leqslant \gamma' < 2\pi$, through the relations $\alpha'=\alpha$, $\beta'=\beta$, $\gamma'=\gamma \bmod 2\pi$.

The Schr{\"o}\-ding\-er equation for this wave function is obtained by quantizing the classical rigid rotator (for which the configuration space is ${\mathbb R}^3 \times SO(3)$). The Pauli equation, which is the standard equation describing non-relativistic spin-1/2 particles, can be obtained from it. Actually, because of the use of the double covering group, this theory contains the Pauli equation twice \cite[pp.\ 439-441]{holland93b}. 

The beables in Holland's model are a position ${\bf x}$ and an orientation ${\boldsymbol \alpha}' \in SO(3)$ in 3-space. The orientation ${\boldsymbol \alpha}'$ is obtained from a beable configuration ${\boldsymbol \alpha} \in SU(2)$. The position ${\bf x}$ represents the center of mass of a rigid homogeneous sphere and ${\boldsymbol \alpha}'$ represents the orientation of a fixed axis through the center of this sphere. The exact form of the Schr{\"o}\-ding\-er equation and the guidance equations are not really important here. The equilibrium distribution for the beables $({\bf x},{\boldsymbol \alpha})$ is given by $|\psi({\bf x},{\boldsymbol \alpha},t)|^2 d^3 x d\Omega$, where $d\Omega= \sin \alpha d\alpha d\beta d\gamma$ is the measure on $SU(2)$. 

This pilot-wave model has the property that any two wave functions are always very much overlapping in ${\boldsymbol \alpha}$-space. This is for example illustrated by considering states of the form:
\begin{equation}
\psi({\bf x},{\boldsymbol \alpha},t) = \psi_+({\bf x},t) u_+({\boldsymbol \alpha}) + \psi_{-}({\bf x},t) u_{-}({\boldsymbol \alpha})\,,
\label{fh.1}
\end{equation}
where
\begin{equation}
u_+ = \frac{1}{2\sqrt{2} \pi} \cos(\alpha/2) e^{-\ii( \beta +\gamma)/2}\,,\quad   u_- = \frac{-\ii}{2\sqrt{2} \pi} \sin(\alpha/2) e^{\ii(  \beta - \gamma)/2}\,.
\label{fh.2}
\end{equation}
The states $u_+$ and $u_-$ respectively correspond to spin up and spin down along the $x_3$-direction and $(\psi_+,\psi_-)$ forms the Pauli spinor. States of this form already contain the full Pauli theory. The fact that different wave functions always have much overlap in ${\boldsymbol \alpha}$-space implies that the beable ${\boldsymbol \alpha}$, or ${\boldsymbol \alpha}'$, never records any outcome of a measurement, in particular it does not record the spin of the system. Of course we are considering only a single particle here and there is no need for measurement results to be recorded in the beables of one particle. On the other hand, if we perform a spin measurement with a Stern-Gerlach device, then the spin will get recorded in the position beable ${\bf x}$, since the wave function will develop non-overlapping terms in position space. For example, if we perform a spin measurement in the $x_3$-direction, then the wave functions $\psi_+({\bf x},t)$ and $\psi_-({\bf x},t)$ will separate in the $x_3$-direction. As such, the position beable ${\bf x}$ will either move up or down in the $x_3$-direction, thereby displaying the outcome of the spin measurement. 

Considering the many particle case, it is unclear to what extent measurement results get recorded in the configuration of angular beables. However, this is not really important if the goal is merely to show that the pilot-wave model reproduces the standard quantum predictions, since measurement results will at least be recorded in the positions of the rigid spheres. 

The insight that measurement results generally get recorded in positions of things, like the positions of measurement needles, led Bell \cite{bell66,bell71} to formulate a more natural pilot-wave model for non-relativistic spin-1/2 particles where the only beables are particle positions (starting from the Pauli equation).

\subsubsection{Fermionic Schr\"odinger field}\label{fermionicschrodinger}
Let us now turn to Holland's pilot-wave model for fermionic fields. Holland illustrated the construction of his model for the case of the fermionically quantized Schr{\"o}\-ding\-er field, but his method applies equally well to other fermionic quantum fields, like for example the quantized Dirac field. 

\paragraph{Functional Schr\"odinger picture}
The fermionically quantized Schr{\"o}\-ding\-er field is obtained by imposing the anti-commutation relations 
\begin{equation}
\{{\widehat \psi}({\bf x}),{\widehat \psi^*}({\bf y})\} = \delta({\bf x}-{\bf y})\,,\quad  \{ \widehat{\psi}({\bf x}),\widehat{\psi}({\bf y})\}=\{ \widehat{\psi}^*({\bf x}),\widehat{\psi}^*({\bf y})\}=0\,. 
\label{fh.0}
\end{equation}
The corresponding Hamiltonian operator has the same form as the bosonic Hamiltonian, which was given in equation (\ref{s.17}) (the term $E_0$ was not present in Holland's presentation). 

Holland chooses a representation for these field operators in terms of angular variables $(\alpha_{\bf k},\beta_{\bf k},\gamma_{\bf k})$ in momentum space.{\footnote{We do not reproduce the equations of Holland's model here. There is however one small correction we want to suggest to Holland's presentation. In order to express the fermionic creation and annihilation operators ${\widehat a}^\dagger_{\bf k}$ and ${\widehat a}_{\bf k}$ in terms of angular momentum operators, a total order relation $\leqslant$ is needed on the set of possible momenta \cite{jordan28}. This order relation can be chosen completely arbitrary. However, the order relation Holland specifies is only a partial one and not a total one, i.e.\ not every two momenta can be compared according to Holland's order relation. Holland's order relation is defined by: ${\bf k}' < {\bf k}$ if $k'_i < k_i$ for one $i=1,2,3$, and $k'_j \leqslant k_j$ for $j \neq i$, which does not yield a total order relation. However a total order relation can easily be found. One could for example consider the one defined by: ${\bf k}' < {\bf k}$ if $k'_i < k_i$ for one $i=1,2,3$, and $k'_j \leqslant k_j$ for $j > i$. However, this correction does not imply any formal change to Holland's presentation. It should just be kept in mind that some particular total order relation should be chosen.\label{totalorder}}} For each momentum ${\bf k}$, the angular variables, collectively denoted as ${\boldsymbol \alpha}_{\bf k}=(\alpha_{\bf k},\beta_{\bf k},\gamma_{\bf k})$, with range $0 \leqslant \alpha_{\bf k} \leqslant \pi$, $0 \leqslant \beta_{\bf k} < 2\pi$, $0 \leqslant \gamma_{\bf k} < 4\pi$, parametrize the group $SU(2)$. Holland further assumes a finite spatial volume so that the set of momenta is discrete.

In this representation the field operators act on wave functionals $\Psi({\boldsymbol \alpha},t)$. The wave functionals can be written as superpositions of states of the form $\prod_{\bf k} u_{{\bf k}n_{\bf k}} ({\boldsymbol \alpha}_{\bf k})$, where each $u_{{\bf k}n_{\bf k}}({\boldsymbol \alpha}_{\bf k})$, with $n_{\bf k}=\pm$, is given by ({\ref{fh.2}}) for each momentum ${\bf k}$. For example, the vacuum state is given by 
\begin{equation}
\Psi_0({\boldsymbol \alpha}) = \prod_{\bf k} u_{{\bf k} -} ({\boldsymbol \alpha}_{\bf k})\,. 
\label{fh.4}
\end{equation}
States describing $n$ particles with definite momenta ${\bf k}_1,\dots,{\bf k}_n$ are given by
\begin{equation}
\Psi_{{\bf k}_1 \dots {\bf k}_n}({\boldsymbol \alpha}) = s({\bf k}_1,\dots,{\bf k}_n) u_{{\bf k}_1 +}  ({\boldsymbol \alpha}_{{\bf k}_1}) \dots u_{{\bf k}_n +}({\boldsymbol \alpha}_{{\bf k}_n}) \prod_{{\bf k} \neq {\bf k}_1, \dots, {\bf k}_n } u_{{\bf k} -} ({\boldsymbol \alpha}_{\bf k})\,,
\label{fh.5}
\end{equation}
where $s({\bf k}_1,\dots,{\bf k}_n)$ is the sign of the permutation that maps $({\bf k}_1,\dots,{\bf k}_n)$ to $({\bf k}'_1,\dots,{\bf k}'_n)$, where $\{{\bf k}_1,\dots,{\bf k}_n\} = \{{\bf k}'_1,\dots,{\bf k}'_n\}$ and ${\bf k}'_1 < \dots < {\bf k}'_n$ (see footnote \ref{totalorder} for more on the order relation on the set of momenta). (More precisely we have $\Psi_{{\bf k}_1 \dots {\bf k}_n}({\boldsymbol \alpha}) = \la {\boldsymbol \alpha} | {\widehat a}^\dagger_{{\bf k}_1} \dots {\widehat a}^\dagger_{{\bf k}_n} |\Psi_0 \ra$, where $ {\widehat a}^\dagger_{{\bf k}}$ are the fermionic creation operators of particles with momentum ${\bf k}$. Using \cite[pp.\ 449-457]{holland93b} and \cite{holland881}, it can easily be verified that this expression reduces to the one in \eqref{fh.5}.)

\paragraph{Ontology}
Holland starts with introducing beables ${\boldsymbol \alpha}_{\bf k}$ for the operators ${\widehat {\boldsymbol \alpha}}_{\bf k}$. The quantum equilibrium distribution for the beables is given by $|\Psi({\boldsymbol \alpha})|^2 \prod_k d \Omega_k$. However, these beables form a field in momentum space and unfortunately Holland does not explicitly specify how these beables relate to some ontology in space.

When discussing the classical limit of his model \cite[pp.\ 455-457]{holland93b}, Holland defines a complex-valued field $\psi({\bf x},t)$ in terms of the beables ${\boldsymbol \alpha}_{\bf k}$ and the wave functional, for the special case of a vanishing quantum potential (in \cite{holland93b} Holland adopts, just as Bohm for the electromagnetic field, a second order view in terms of the quantum potential). Possibly a similar definition can be given in general (i.e.\ also when the quantum potential does not vanish). One way to do this could be by using Holland's local expectation value for the operator ${\widehat \psi}({\bf x})$.{\footnote{Holland only introduced local expectation values for Hermitian operators \cite{holland93b}. One could define the local expectation value for ${\widehat \psi}({\bf x})$ as the linear combination of the local expectation values for the Hermitian operators ${\widehat \psi}_r({\bf x}) = ({\widehat \psi}({\bf x}) + {\widehat \psi}^\dagger({\bf x}) )/2$ and ${\widehat \psi}_i({\bf x}) = -\ii ({\widehat \psi}({\bf x}) - {\widehat \psi}^\dagger({\bf x}) )/2$. Note that the local expectation value for ${\widehat \psi}({\bf x})$ would be a complex-valued field and hence would not satisfy the anti-commutation relations \eqref{fh.0} of the field operator ${\widehat \psi}({\bf x})$.}}

Another possibility would be to introduce a beable for the particle number density operator ${\widehat \psi}^\dagger({\bf x}){\widehat \psi}({\bf x})$, again by means of Holland's local expectation value.

\paragraph{Reproducing the quantum predictions}
Whether or not Holland's model is capable of reproducing the standard quantum predictions depends of course on the particular choice of ontology in physical space. Macroscopically different states should correspond to different beable configurations. Therefore it is necessary (at least for the type of beable suggested above) that the wave functionals are non-overlapping in the configuration space of angular beables. This guarantees that macroscopically different states correspond to different beable configurations ${\boldsymbol \alpha}$. It is then further necessary that the different beable configurations ${\boldsymbol \alpha}$ correspond to different beables in physical space. Since our examples of possibly ontologies in physical space are rather complicated constructions involving the wave functional $\Psi({\boldsymbol \alpha})$, it is not immediately clear that this is guaranteed. 

It is easy to see that states containing only a small number of particles have much overlap. However some states have little overlap. Consider for example the state $\Psi_{{\bf k}_1 \dots {\bf k}_n}({\boldsymbol \alpha})$. This state will have little overlap with the vacuum state $\Psi_0({\boldsymbol \alpha})$ in the case the number of particles $n$ is high enough. This follows from the fact that the overlap of the functions $u_{{\bf k}_1 +}({\boldsymbol \alpha}_{{\bf k}_1}) \dots u_{{\bf k}_n +}({\boldsymbol \alpha}_{{\bf k}_n})$ and $u_{{\bf k}_1 -}({\boldsymbol \alpha}_{{\bf k}_1}) \dots u_{{\bf k}_n -}({\boldsymbol \alpha}_{{\bf k}_n})$ decreases when $n$ increases (this can be seen by considering the law of large numbers{\footnote{I am grateful to Owen Maroney for this observation.}}). Similarly, two states $\Psi_{{\bf k}_1 \dots {\bf k}_n}({\boldsymbol \alpha})$ and $\Psi_{{\bf k}'_1 \dots {\bf k}'_{n'}}({\boldsymbol \alpha})$ will have negligible overlap when the number of momenta in the symmetric difference of the sets $\{ {\bf k}_1, \dots, {\bf k}_n\}$ and $\{ {\bf k}_1, \dots, {\bf k}_n\}$ is sufficiently large. It is an open question whether states describing localized macroscopic systems have negligible overlap.

So it is as yet unclear whether Holland's model is capable of reproducing the predictions of standard quantum theory. In the following section an alternative but similar model is considered for which this question is much easier to analyse. However, the upshot of that analysis is that the model seems inadequate.

\subsubsection{Alternative model for the fermionic Schr\"odinger field}\label{alternativemodel}
Since it is rather hard to analyse the viability of Holland's model for the fermionic field due to its complicated ontology, it is worthwhile to consider the following simpler alternative. In developing his pilot-wave model, Holland started with the field operators ${\widehat a}_{\bf k}$ and ${\widehat a}^\dagger_{\bf k}$ in momentum space, which satisfy the anti-commutation relations $\{{\widehat a}_{\bf k},{\widehat a}^\dagger_{{\bf k}'}\} = \delta_{{\bf k}{\bf k}'}$. By using a representation in terms of angular momentum operators in momentum space, he was led to introduce the beables ${\boldsymbol \alpha}_{\bf k}$. Instead of starting with the field operators in momentum space, one could also start with considering operators in physical 3-space. Assuming a spatial lattice, instead of a continuum space, the anti-commutation relations for the field operators ${\widehat \psi}_{\bf x}$ and ${\widehat \psi}^*_{\bf x}$ read $\{{\widehat \psi}_{\bf x},{\widehat \psi^*}_{{\bf x}'}\} = \delta_{{\bf x}{\bf x}'}$, where ${\bf x}$ and ${\bf x}'$ are lattice sites. These anti-commutation relations are formally the same as the anti-commutation relations for the field operators in momentum space.{\footnote{Holland carried out his analysis for a discrete set of momenta. However, probably a similar (formal) analysis can be carried out in the case of a continuous set of momenta. In that case we could, equivalently, assume a physical space which is continuous instead of discrete.}} One could therefore adopt the techniques introduced by Holland and use a representation of these anti-commutation relations in terms of angular momentum operators. This time the angular momentum operators would be operators in physical space. This would result in beables ${\boldsymbol \alpha}_{\bf x}=(\alpha_{\bf x},\beta_{\bf x},\gamma_{\bf x})$ in physical space. While one could further assume a rigid body at each lattice point ${\bf x}$, whose rotational motion is described by the angles ${\boldsymbol \alpha}_{\bf x}$ (similarly as in Holland's model for non-relativistic spin-1/2 particles), there is no need for that. 

We do not present the dynamics of the beables here. It could easily be found by using the results in \cite{struyve09a}. Even without knowing the dynamics explicitly we can say something about the viability of the model. This could be done by comparing the typical beable configurations corresponding to a state $\Psi_M$ which represents matter in a certain region $V$ in physical space and a state $\Psi_e$ which represents an empty region $V$. We have, in analogy with \eqref{fh.4} and \eqref{fh.5}, that 
\begin{equation}
\Psi_e({\boldsymbol \alpha}) = \prod_{{\bf x} {\textrm{ in }}V}  u_{{\bf x} -} ({\boldsymbol \alpha}_{\bf x}) 
\label{fh.7}
\end{equation}
and 
\begin{equation}
\Psi_M({\boldsymbol \alpha}) = s({\bf x}_1,\dots,{\bf x}_n) u_{{\bf x}_1 +}  ({\boldsymbol \alpha}_{{\bf x}_1}) \dots u_{{\bf x}_n +}({\boldsymbol \alpha}_{{\bf x}_n}) \prod_{\substack{{\bf x} {\textrm{ in }}V \\ {\bf x} \neq {\bf x}_1, \dots, {\bf x}_n  } } u_{{\bf x} -} ({\boldsymbol \alpha}_{\bf x})\,,
\label{fh.6}
\end{equation}
where ${\bf x}_1,\dots,{\bf x}_n$ are the positions of the particles in $V$. For our purposes the environment can be ignored in the states. 

Note that matter will never be distinguished from empty space from the beables $\beta_{\bf x}$ or $\gamma_{\bf x}$, because the states states $\Psi_M$ and $\Psi_e$ yield the same distribution for them. They could only be distinguished from the beables $\alpha_{\bf x}$. This could be done by considering the quantity $A_V= \sum_{{\bf x} {\textrm{ in }}V} \alpha_{\bf x} / n$. If this quantity is typically different for the states $\Psi_M$ and $\Psi_e$ then one could read off from the beables whether the region $V$ is filled with matter or not. For a state $\Psi$ the expected value for the quantity $A_V$ is given by $\langle A_V \rangle_\Psi = \int  A_V |\Psi|^2\prod_{{\bf x} {\textrm{ in }}V } d \Omega_{\bf x}$. Matter can be distinguished from empty space from the beables if 
\begin{equation}
\left| \langle A_V \rangle_{\Psi_M} - \langle A_V \rangle_{\Psi_e} \right| \gg {\textrm{max}} \left( \Delta_{\Psi_M}A_V ,\Delta_{\Psi_e} A_V  \right)\,,
\label{fh.8}
\end{equation}
where $\Delta_{\Psi}A_V = \sqrt{\langle A_V^2 \rangle_\Psi - \langle A_V \rangle_\Psi^2}$ is the standard deviation of $A_V$ for the state $\Psi$. For a cubic region $V$ with edges of length $L$, matter with particle number density $\rho$ and a lattice spacing $a$, we find in Appendix \ref{appendixb} that \eqref{fh.8} reduces to
\begin{equation}
La\rho^{2/3} \gg 1 \,.
\label{fh.9}
\end{equation}
Assume one particle per atomic distance cubed, so that $\rho = 10^{30}/{\textrm m}^3$. Then \eqref{fh.9} becomes $La  \gg 10^{-20} {\textrm m}^2$. If we further assume a lattice spacing of the order of the Planck length, i.e.\ $a = 10^{-35}{\textrm m}$, then $L \gg 10^{15}{\textrm m}$. This means that in this case we can only distinguish matter from empty space by considering length scales that are much bigger then $10^{15}{\textrm m}$. This is clearly unreasonable. The lattice spacing could of course be taken much larger than the Planck length in a non-relativistic theory. But presumably it should not be taken much higher then $10^{-15}{\textrm m}$ which is only three orders lower than the Compton wavelength of the electron. For the latter value of the lattice spacing, $L$ must be much bigger then $10^{-5}{\textrm m}$. This is still a rather large distance, as we can easily distinguish such length scales with the naked eye (the diameter of a human hair typically ranges from $10^{-5}{\textrm m}$ to $10^{-4}{\textrm m}$). This implies that it seems very unlikely that the beables in this model yield an image of general macroscopic matter distributions. As such the model seems to fail to reproduce the standard quantum predictions. 

There also seems to be little hope that an application of this model to relativistic quantum field theory will yield a better result. In a relativistic model the beables would consist of four set of angles at each point in space (one corresponding to each component of the Dirac spinor) while the lattice spacing should be taken much smaller.

\subsection{Valentini's approach: Grassmann fields as beables}\label{valentinismodelfermions}
In \cite{valentini92,valentini96} Valentini suggested a pilot-wave model for fermionic fields in which the beables are anti-commuting fields, also called {\em Grassmann fields}. In this section we point out a problem with this approach. We will see that although it is possible to introduce dynamics for anti-commuting fields, it is unclear what the actual guidance equations and corresponding equilibrium distribution should be, in order to have a pilot-wave model that reproduces the standard quantum predictions.

The starting point of Valentini's approach is to write the fermionic field theory in a functional Schr{\"o}\-ding\-er picture in which the quantum states are functionals defined on a space of anti-commuting fields. Although this functional Schr{\"o}\-ding\-er picture, which was initiated by Floreanini and Jackiw \cite{floreanini88}, bears some resemblance to the functional Schr{\"o}\-ding\-er picture for bosonic field theories, there are some important differences. One is that the fermionic wave functionals are not complex-valued, but take values in a Grassmann algebra. A consequence is that the wave functional does not yield a transition amplitude for finding the state in a certain field configuration. 

In order to point out clearly the problem with Valentini's model, we present the functional Schr{\"o}\-ding\-er picture for fermionic fields in some detail in section \ref{functschgrassmann}. The reader who is familiar with this picture can immediately move to section \ref{pwmgrassmann}. 

As mentioned in section \ref{valentinisway}, Valentini presented his pilot-wave model for the quantized Van der Waerden field. Valentini used the Van der Waerden theory, instead of the equivalent Dirac theory, in order to circumvent the problem of the constraints in the Dirac theory. For simplicity we do not discuss Valentini's pilot-wave model for the quantized Van der Waerden field, but for the fermionically quantized Schr{\"o}\-ding\-er field (which was already introduced in section \ref{fermionicschrodinger}). The formulation of the functional Schr{\"o}\-ding\-er picture for the quantized Schr{\"o}\-ding\-er field is completely similar to that for the quantized Van der Waerden field or the quantized Dirac field. The functional Schr{\"o}\-ding\-er picture for the Dirac field can for example be found in \cite{kiefer94}.

\subsubsection{Functional Schr{\"o}\-ding\-er picture in terms of Grassmann fields}\label{functschgrassmann}
In order to set up the functional Schr{\"o}\-ding\-er picture for the fermionically quantized Schr{\"o}\-ding\-er field, we first introduce some basic elements of Grassmann algebras and analysis over these algebras. Details can be found in \cite{berezin66,berezin87,dewitt92}.

Consider two fields $w({\bf x})$ and $w^{\dagger}({\bf x})$ which satisfy 
\begin{equation}
\{w({\bf x}),w({\bf y}) \} =\{w({\bf x}),w^{\dagger}({\bf y})\}=\{w^{\dagger}({\bf x}),w^{\dagger}({\bf y})\}=0\,.
\label{fv.1}
\end{equation} 
These fields are {\em Grassmann fields} (i.e.\ for each ${\bf x}$ the values $w({\bf x})$ and $w^{\dagger}({\bf x})$ are {\em Grassmann numbers}). The algebra generated by these fields forms an infinite dimensional {\em Grassmann algebra}, which we denote by $\Lambda$. Its elements are functionals $G(w,w^{\dagger})$ of the form
\begin{eqnarray}
G &=& G_{00} + \int d^3 x_1 G_{10} ({\bf x}_1) w({\bf x}_1) +  \int d^3 x_1 G_{01} ({\bf x}_1) w^\dagger({\bf x}_1) + \dots \nonumber\\
&& \mbox{} + \int d^3 x_1 \dots d^3 x_{n+m} G_{nm} ({\bf x}_1,\dots,{\bf x}_{n+m})  w({\bf x}_1) \dots w({\bf x}_n)  w^\dagger({\bf x}_{n+1}) \dots  w^\dagger({\bf x}_{n+m}) \nonumber\\
&&\mbox{}  +\dots \,,
\label{fv.2}
\end{eqnarray}
where $G_{00}$ is a complex number and where $G_{nm}({\bf x}_1,\dots,{\bf x}_{n+m})$, with $n+m>0$, are completely anti-symmetric complex-valued distributions. There is an involutive map $G \mapsto G^{\dagger}$ on $\Lambda$, called {\em complex conjugation}, which is linear under addition, anti-linear under multiplication by complex numbers, satisfies $(G_1G_2)^\dagger = G^\dagger_2 G^\dagger_1$ and $\left(w({\bf x})\right)^\dagger = w^\dagger({\bf x})$, and which acts as the ordinary complex conjugation on complex numbers. An element $G$ of $\Lambda$ is called {\em odd} if $G_{nm}=0$ for all $(m,n)$ for which $n+m$ is even. All odd elements anti-commute among themselves. The subspace of odd elements is denoted by $\Lambda_o$. 

We can now introduce the space ${\mathcal{F}}$ of $\Lambda$-valued functionals $F(u,u^\dagger)$, which are defined on the space of $\Lambda_o$-valued fields $u({\bf x})$ and $u^\dagger({\bf x})$,{\footnote{As is standard practice, we treat $u({\bf x})$ and $u^\dagger({\bf x})$ as independent fields. Alternatively one could consider two real fields $u_1({\bf x})$ and $u_2({\bf x})$ (the fact that they are real means that $u_i({\bf x})=u^\dagger_i({\bf x})$, $i=1,2$), and define $u({\bf x})=(u_1({\bf x}) + \ii u_2({\bf x}))/\sqrt{2}$ and $u^\dagger({\bf x})=(u_1({\bf x}) - \ii u_2({\bf x}))/\sqrt{2}$.}} and which are of the form{\footnote{One can also define $\Lambda_o$-valued fields that are not of the form \eqref{fv.3} by considering the $F_{nm}$ to be $\Lambda$-valued instead of complex-valued. However such functionals are of no interest here.}}
\begin{eqnarray}
F &=& F_{00} + \int d^3 x_1 F_{10} ({\bf x}_1) u({\bf x}_1) +  \int d^3 x_1 F_{01} ({\bf x}_1) u^\dagger({\bf x}_1) + \dots \nonumber\\
&& \mbox{} + \int d^3 x_1 \dots d^3 x_{n+m} F_{nm} ({\bf x}_1,\dots,{\bf x}_{n+m})  u({\bf x}_1) \dots u({\bf x}_n)  u^\dagger({\bf x}_{n+1}) \dots  u^\dagger({\bf x}_{n+m}) \nonumber\\
&&\mbox{}  +\dots \,,
\label{fv.3}
\end{eqnarray}
where $F_{00}$ is a complex number and where $F_{nm}({\bf x}_1,\dots,{\bf x}_{n+m})$, with $n+m>0$, are completely anti-symmetric complex-valued distributions. 

For functionals in ${\mathcal{F}}$ the functional derivatives ${\delta F(u,u^\dagger)}/{\delta u({\bf x})}$ and ${\delta F(u,u^\dagger)}/{\delta u({\bf x})}$ exist for all $\Lambda_o$-valued fields $u({\bf x})$ and $u^\dagger({\bf x})$. The functional derivative ${\delta F(u,u^\dagger)}/{\delta u({\bf x})}$ is hereby defined as the coefficient that arises in the variation $\delta F(u,u^\dagger)$ under an infinitesimal variation $\delta u({\bf x})$ of $u({\bf x})$ in the space of $\Lambda_o$-valued fields \cite{dewitt92}, i.e.
\begin{equation}
\delta F(u,u^\dagger) = \int d^3x \delta u({\bf x}) \frac{\delta F(u,u^\dagger)}{\delta u({\bf x})} \,.
\label{fv.4}
\end{equation}
The functional derivative ${\delta F(u,u^\dagger)}/{\delta u^\dagger({\bf x})}$ is defined in a similar way. These definitions imply that the functional derivatives anti-commute among each other and that they satisfy the properties
\begin{equation}
\left\{u({\bf x}),\frac{\delta}{\delta u({\bf y})}\right\} = \left\{u^\dagger({\bf x}),\frac{\delta}{\delta u^\dagger({\bf y})}\right\} = \delta({\bf x}-{\bf y})\,,
\label{fv.5}
\end{equation}
\begin{equation}
 \left\{u({\bf x}),\frac{\delta}{\delta u^\dagger({\bf y})}\right\} = \left\{u^\dagger({\bf x}),\frac{\delta}{\delta u({\bf y})}\right\} = 0 \,.
\label{fv.6}
\end{equation}

For functionals in ${\mathcal{F}}$ one can also define functional integrals. First of all, for a fixed point ${\bf x}$, the integral over $u({\bf x})$ is defined by the relations $\int du({\bf x}) =0$ and $\int du({\bf x}) u({\bf x}) =1 $. The integration is hence defined formally, i.e.\ measure-theoretical concepts play no role. The functional integral $\int {\mathcal{D}}u F(u,u^\dagger)$ is then defined as the (formal) product integral $\int (\prod_{{\bf x}} d u({\bf x})) F(u,u^\dagger)$. The functional integral $\int {\mathcal{D}}u^\dagger F(u,u^\dagger)$ is defined similarly.

In the functional Schr{\"o}\-ding\-er picture for the quantized Schr{\"o}\-ding\-er field, the field operators $\widehat{\psi}({\bf x})$ and $\widehat{\psi}^*({\bf x})$ are realized by the representation{\footnote{Just as in bosonic field theory, there exist various other possible representations of the field operators. Another possible representation is $\widehat{\psi}({\bf x}) \to  u({\bf x})$, $\widehat{\psi}^*({\bf x}) \to  \delta / \delta  u^\dagger({\bf x})$,  see e.g.\ \cite[pp.\ 217-223]{hatfield91} or \cite{hallin95}.}}
\begin{equation}
\widehat{\psi}({\bf x}) \to \frac{1}{\sqrt{2}} \left( u({\bf x}) + \frac{ \delta }{ \delta  u^\dagger({\bf x})} \right)\,,\qquad \widehat{\psi}^*({\bf x}) \to \frac{1}{\sqrt{2}}  \left( u^\dagger({\bf x}) + \frac{ \delta }{\delta u({\bf x})}\right)\,,
\label{fv.9}
\end{equation}
which act on quantum states which are elements of ${\mathcal{F}}$. The associated inner product is defined as
\begin{equation}
\langle \Psi_1 | \Psi_2 \rangle = \int \mathcal{D}u^{\dagger} \mathcal{D} u \Psi^*_1 \Psi_2 \,,
\label{fv.7}
\end{equation}
where $\Psi^*$ is the dual of $\Psi$ given by
\begin{equation}
\Psi^*(u,u^{\dagger}) = \int \mathcal{D}{\bar u}^{\dagger} \mathcal{D} {\bar u} \exp({\bar u}u^{\dagger}+ {\bar u}^{\dagger} u) \Psi^{\dagger}({\bar u},{\bar u}^{\dagger})\,,
\label{fv.8}
\end{equation}
where we used the notation ${\bar u} u^{\dagger} = \int d^3 x {\bar u}({\bf x})  u^{\dagger}({\bf x})$ and similarly for ${\bar u}^{\dagger} u$ (the integration variables ${\bar u}({\bf x})$ and ${\bar u}^{\dagger}({\bf x})$ are $\Lambda_0$-valued fields in space). The functional Schr{\"o}\-ding\-er equation is then obtained by using this representation for the Hamiltonian operator (which has the same form as the one for the bosonically quantized Schr{\"o}\-ding\-er field, cf.\ (\ref{s.17})). For the functional Schr{\"o}\-ding\-er representation for the quantized Dirac field one just needs to add a spinor index $a=1,\dots,4$ to the Grassmann fields.

\subsubsection{Pilot-wave model}\label{pwmgrassmann}
Let us now turn to the pilot-wave model of Valentini. Valentini postulated anti-com\-mu\-ting, i.e.\ $\Lambda_o$-valued, field beables $u({\bf x})$ and $u^{\dagger}({\bf x})$, which are guided by the wave functional $\Psi(u,u^\dagger,t)$ through guidance equations which are of the form ${\dot u} = v^{\Psi}(u,u^\dagger,t)$ and ${\dot u}^\dagger = (v^{\Psi}(u,u^\dagger,t))^\dagger$. The exact form of the guidance equations is not really important here. But we do want to note that there are actually some technical problems with those guidance equations. They for example contain inverses with respect to the multiplication of elements of the Grassmann algebra, but these are not always defined. However, these problems, which were discussed in detail in \cite{struyve05}, are of minor importance, since, in principle, there is no problem for having equations of motion for anti-commuting fields (see for example also the generalization of classical mechanics in which some of the dynamical variables have anti-commuting values \cite{martin59,berezin77,casalbuoni761,casalbuoni762}). 

The more important problem is that there seems to be no real basis for adopting Valentini's guidance equations. Valentini regarded the quantity $\Psi^\dagger(u,u^\dagger,t) \Psi(u,u^\dagger,t)$ as the density of fields and found the guidance equations by considering the asociated continuity equation.{\footnote{Note that Valentini regarded $\Psi^\dagger \Psi$ as the density of fields and not $\Psi^* \Psi$. Though, regarding $\Psi^*\Psi$ as the density would be as problematic.}} However, this interpretation of $\Psi^\dagger \Psi$ seems to be unfounded. The quantity $\Psi^\dagger(u,u^\dagger,t) \Psi(u,u^\dagger,t)$ is, just as the wave functional $\Psi(u,u^\dagger,t)$, $\Lambda$-valued.{\footnote{The wave functional $\Psi(u,u^\dagger,t)$ should not be confused with a transition amplitude $\langle u,u^\dagger | \Psi(t) \rangle$: $\Psi(u,u^\dagger,t)$ is an element of the Grassmann algebra $\Lambda$ and $\langle u,u^\dagger | \Psi(t) \rangle$ should be a complex number by the definition of the inner product.}} Moreover, it is unclear what measure should be considered on the space of anti-commuting fields. To our knowledge suitable measure-theoretical notions have not yet been introduced for such a space. Remember also that the integration theory was defined formally and was not based on measure-theoretical notions. Therefore $\Psi^\dagger \Psi$ seems to have no meaning as a density of fields. To be sure, this problem is not a consequence of dealing with a field theory; it would still be present in a regularized theory that describes only a finite number of degrees of freedom. So, in summary, it is unclear what the dynamics and corresponding equilibrium distribution should be so that the standard quantum predictions are reproduced.

\section{A minimalist pilot-wave model}\label{minimalist}
Motivated by the difficulties of introducing field beables for fermionic field theories, Struyve and Westman \cite{struyve06,struyve07c} proposed a model in which there are only beables for the bosonic degrees of freedom and no beables for the fermionic degrees of freedom. This approach was illustrated for quantum electrodynamics. In that case the beable is a transverse field $A^T_i$, representing the transverse part of the vector potential, just as in Bohm's model for the free electromagnetic field. The dynamics is very similar to that of Bohm's model. The difference is that in this case the wave functional also contains fermionic degrees of freedom, which have an effect on the dynamics of the beable $A^T_i$. This makes the beable $A^T_i$ evolve in such a way as to give the appearance of matter.  

It was argued in detail that this model is actually capable of reproducing the predictions of standard quantum field theory, as one can find an image of the visible world, and in particular of measurement outcomes, in the field beable $A^T_i$. However, while this model can arguably be maintained as a logical possibility, it is rather far removed from our everyday experience of the world. 

After presenting this radically minimalist model, Struyve and Westman also illustrated how one could introduce additional beables for the fermionic degrees of freedom \cite{struyve07c}. By means of Holland's local expectation value, beables could be introduced for operators such as the mass density operator or charge density operator. In this way the beables for the fermionic degrees of freedom are completely determined by the field beable $A^T_i$ and the wave functional (while they do not influence the field beable $A^T_i$ or the wave functional).

The beables corresponding to these density operators would behave in a way similar to that of the mass density in collapse theories \cite{bassi03}. For example, in a double-slit experiment with an electron, the mass or charge density of the electron would evolve through both slits and would display some interference pattern just before arriving to the detection screen, only to ``collapse'' to an approximately localized density when an actual position measurement is performed. This collapse happens vecause during this measurement process the wave function of the electron gets entangled with that of a macroscopic pointer and with that of the radiation being emitted and scattering off the macroscopic pointer. The entanglement with the state of radiation is ultimately responsible for the collapse of the mass or charge density of the electron.

\section{Conclusions and outlook}
In this paper we have studied the construction of pilot-wave models for quantum field theory for which the beables are fields. We started with presenting a recipe according to which such pilot-wave models could be developed, but only for bosonic fields. For gauge theories the recipe implies that beables are only introduced for gauge independent degrees of freedom. This recipe can easily be applied in the case of simple quantum field theories, as was illustrated for example for the Schr{\"o}\-ding\-er field and for scalar quantum electrodynamics. For more involved theories, like Yang-Mills theories, this is more complicated and has not yet been achieved. 

In the paper, we did not consider alternative approaches to gauge theories, such as for example loop-based approaches. In a loop-based approach the basic variables are not the gauge potentials, but loops in space or space-time \cite{gambini96,rovelli04,thiemann07}. For example, the free electromagnetic field is easily reformulated in terms of such variables \cite{gambini96}. Considering that such an approach can also be applied to quantum gravity, namely in the loop quantum gravity approach, it might be worthwhile to try to develop a pilot-wave theory in terms of such variables. 

For fermionic quantum field theories it is much harder to construct a pilot-wave model with fields as beables. We considered a model by Holland and a model by Valentini. In Holland's model the beables form a field of angular variables. Further study of this model is required in order to see whether it reproduces the quantum predictions. Valentini's suggestion on the other hand is to introduce anti-commuting fields as beables. However, this suggestion meets the problem of identifying a suitable equilibrium probability distribution over these anti-commuting fields. Without such a distribution it is unclear how to start developing a pilot-wave model.

One model that has field beables for fermionic fields and that is capable of reproducing the standard quantum predictions is the model by Struyve and Westman, which was only briefly discussed here. In this model the beables for the fermionic fields are defined in terms of the beables for the bosonic degrees of freedom and the wave function. 

On the other hand, taking particle positions as beables seems more succesful for fermionic field theories than for bosonic field theories \cite{bell87b,durr02,durr031,durr032,tumulka03,durr04,colin031,colin032,colin033,colin07}. In view of this, it might be worthwile to consider a supersymmetric extension of standard quantum field theory. In a supersymmetric quantum field theory there is a symmetry between bosons and fermions. Maybe this symmetry allows for a more unified approach for introducing beables.

\section{Acknowledgements}
I am very grateful to Antony Valentini for providing me with a draft of his book \cite{valentini09} which contains extensive discussions on field beables from which I benefited a lot. I am further grateful to Jean Bricmont, Samuel Colin, and in particular to Antony Valentini and Hans Westman for valuable discussions, and to Samuel Colin, Sheldon Goldstein, Peter Holland, Roderich Tumulka, Antony Valentini, Hans Westman and two anonymous referees for many valuable comments on the manuscript. 

Most of this work was carried out at the Perimeter Institute for Theoretical Physics. Research there is supported by the Government of Canada through Industry Canada and by the Province of Ontario through the Ministry of Research \& Innovation. Currently the support of the FWO-Flanders is acknowledged.


\appendix
\section{On deriving the guidance equation from analogy with Hamilton-Jacobi theory}\label{appendixa}
The general method to find the guidance equation for some configuration $q$ is to consider the continuity equation for the quantum mechanical probability density $|\psi(q,t)|^2$ and to define the velocity field $v^\psi$ for $q$ as $v^\psi=j^\psi/|\psi|^2$, where $j^\psi$ is the probability current associated to the density $|\psi|^2$.

For many quantum theories of interest that can be obtained by quantizing a classical Lagrangian $L(q,{\dot q})$, it is so that the guidance equations ${\dot q}_n = j^\psi_n/|\psi|^2$ are equivalent to the equations
\begin{equation}
p_n = \frac{\partial S }{ \partial q_n},
\label{a.0}
\end{equation}
where $p_n=\partial L(q,{\dot q})/ \partial q_n$ are the classical momenta and where $S$ is the phase of the wave function, i.e.\ $\psi=|\psi|\exp(\ii S)$ \cite{valentini96}. The latter prescription \eqref{a.0} is inspired by the Hamilton-Jacobi formulation of classical mechanics.

This is for example so for the non-relativistic pilot-wave theory of \db\ and Bohm that was discussed in section \ref{examplepilotwaveinterpretationnonrelativistic}. The guidance equations $\dot{ {\bf x}}_k = m^{-1}_k {\boldsymbol{\nabla}}_k S$ are readily obtained by equating the classical momenta, which are given by ${\bf p}_k = m_k {\dot {\bf x}}_k$, with ${\boldsymbol{\nabla}}_k S$. The pilot-wave theory further yields an equation that resembles the Hamilton-Jacobi equation of classical mechanics. However, this resemblance is merely formal. The differences are discussed in detail in \cite[pp.\ 131-134]{holland93b}.
  
However the prescription \eqref{a.0} to find the guidance equation does not hold in general. We already encountered one example where it doesn't hold, namely in the case of the Dirac field, in section \ref{valentiniswaydiracfield}. However, the Dirac theory involves both fermionic quantization and a singular Lagrangian and it might perhaps be thought that the failure of the prescription is due to either of these facts. However, below we consider another theory which involves bosonic quantization and a non-singular Lagrangian for which the prescription fails too.

The example we have in mind corresponds to the classical Hamiltonian 
\begin{equation}
H=\al_1 p^4 + \al_2 p^2\,,
\label{a.1}
\end{equation}
where $\al_1$ and $\al_2$ are strictly positive real constants. The Lagrangian can be obtained by performing the Legendre transform, with velocity defined by 
\begin{equation}
{\dot q} = \frac{\pa H}{\pa p} = 4 \al_1 p^3 + 2 \al_2 p\,.
\label{a.2}
\end{equation}
Since $\pa^2 H/\pa p^2 = 12 \al_1 p^2 + 2 \al_2 > 0$, the relation above can be inverted to yield the momentum $p$ in terms of the velocity ${\dot q}$. The Lagrangian is then given by $L={\dot q}p - H$, where $p$ is expressed in terms of ${\dot q}$. For the subsequent discussion there is no need to give the Lagrangian explicitly.

Hamilton's equation \eqref{a.2} is equivalent to the defining relation of the momentum in terms of the Lagrangian, i.e.\ $p = \pa L/ \pa {\dot q}$. So the above prescription \eqref{a.0} to find the guidance equation, yields
\begin{equation}
{\dot q} =  4 \al_1 \left(\frac{\pa S}{\pa q} \right)^3 + 2 \al_2 \frac{\pa S}{\pa q} \,.
\label{a.3}
\end{equation}

On the other hand, the quantization of the Hamiltonian gives the Schr\"odinger equation:
\begin{equation}
\ii \frac{\pa \psi(q,t)}{\pa t} = \left( \al_1 \frac{\pa^4}{\pa q^4} -  \al_2 \frac{\pa^2}{\pa q^2} \right) \psi(q,t) \,,
\label{a.4}
\end{equation}
for which the corresponding continuity equation reads
\begin{equation}
\frac{\pa |\psi|^2}{\pa t} +  \frac{\pa j^\psi}{\pa q} = 0\,,
\label{a.5}
\end{equation}
with
\begin{eqnarray}
j^\psi &=& 2 \textrm{Im} \left[ \al_1 \left( \frac{\pa^3 \psi^*}{\pa q^3} \psi + \frac{\pa \psi^*}{\pa q} \frac{\pa^2 \psi}{\pa q^2} \right) + \al_2 \psi^*  \frac{\pa \psi}{\pa q} \right] \,,\nonumber\\
&=& 4 \al_1 \left[  R^2 \left( \frac{\pa S}{\pa q} \right)^3 - 2R  \frac{\pa^2 R}{\pa q^2}  \frac{\pa S}{\pa q} + \left( \frac{\pa R}{\pa q} \right)^2 \frac{\pa S}{\pa q}   -     R\frac{\pa R}{\pa q}  \frac{\pa^2 S}{\pa q^2}  - \frac{1}{2} R^2 \frac{\pa^3 S}{\pa q^3} \right] \nonumber\\
&& \mbox{} + 2 \al_2 R^2 \frac{\pa S}{\pa q}\,,
\label{a.6}
\end{eqnarray}
where we used the polar decomposition $\psi=Re^{\ii S}$ in the last line. The resulting guidance equation reads
\begin{eqnarray}
{\dot q}  &=& \frac{j^\psi}{|\psi|^2} \nonumber\\
&=& 4 \al_1 \left[  \left( \frac{\pa S}{\pa q} \right)^3 - 2R^{-1}  \frac{\pa^2 R}{\pa q^2}  \frac{\pa S}{\pa q} + R^{-2} \left( \frac{\pa R}{\pa q} \right)^2 \frac{\pa S}{\pa q}   -     R^{-1} \frac{\pa R}{\pa q}  \frac{\pa^2 S}{\pa q^2}  - \frac{1}{2} \frac{\pa^3 S}{\pa q^3} \right] \nonumber\\
&& \mbox{}  + 2 \al_2 \frac{\pa S}{\pa q} \,.
\label{a.7}
\end{eqnarray}
Hence we see that the correct guidance equation, namely \eqref{a.7}, contains extra terms not present in \eqref{a.3}.

\section{Evaluation of the expression \eqref{fh.8}}\label{appendixb}
With
\begin{equation}
\langle \alpha \rangle_{u_+} = \int d \Omega \alpha |u_+|^2 = \frac{5}{8} \pi \,, \qquad \langle \alpha \rangle_{u_-} = \int d \Omega \alpha |u_-|^2 =\frac{3}{8} \pi \,,
\label{b.1}
\end{equation}
\begin{equation}
\Delta_{u_+} \alpha = \sqrt{\langle \alpha^2 \rangle_{u_+} -  \langle \alpha \rangle_{u_+}^2}= \sqrt{\frac{15 \pi^2}{64} - 2}\,,
\label{b.2}
\end{equation}
\begin{equation}
\Delta_{u_-} \alpha = \sqrt{\langle \alpha^2 \rangle_{u_-} -  \langle \alpha \rangle_{u_-}^2}= \sqrt{\frac{15 \pi^2}{64} - 2}\,,
\label{b.3}
\end{equation}
$n$ the number of particles in $V$, and $n_l$ the number of lattice sites in $V$, we have
\begin{equation}
\langle A_V \rangle_{\Psi_M} = \frac{1}{n_l} \left( n \langle \alpha \rangle_{u_+} + (n_l-n)\langle \alpha \rangle_{u_-} \right) \,, \qquad  \langle A_V \rangle_{\Psi_e} = \langle \alpha \rangle_{u_-} \,,
\label{b.4}
\end{equation}
\begin{equation}
\Delta_{\Psi_M}A_V  = \frac{1}{\sqrt{n_l}} \Delta_{u_+} \alpha = \frac{1}{\sqrt{n_l}} \Delta_{u_-} \alpha =\Delta_{\Psi_e}A_V \,,
\label{b.5}
\end{equation}
so that \eqref{fh.8} reduces to 
\begin{equation}
\frac{n}{\sqrt{n_l}} \gg \frac{\Delta_{u_-} \alpha}{\langle \alpha \rangle_{u_+} - \langle \alpha \rangle_{u_-}} = \frac{4}{\pi} \sqrt{\frac{15 \pi^2}{64} - 2} \,.
\label{b.6}
\end{equation}
With $a$ the lattice spacing and $\rho$ the particle number density of the matter under consideration, and assuming a cubic volume $V$ with edge of length $L$, we have $n=L^3 \rho$ and $n_l = (L/a)^3$. Since the right hand side of \eqref{b.6} is of the order one, \eqref{b.6} reduces to
\begin{equation}
La\rho^{2/3} \gg 1 \,.
\label{b.7}
\end{equation}

\end{document}